\documentclass[12pt,a4paper]{article}

\usepackage[margin=1in]{geometry} 
\usepackage{placeins}  
\usepackage{booktabs}
\usepackage{threeparttable}
\usepackage[english]{babel}
\usepackage{graphicx}
\usepackage{amsfonts}
\usepackage{amsmath}
\usepackage{amsthm}
\usepackage{amssymb}
\usepackage{mathrsfs}
\usepackage{mathtools, amssymb}
\usepackage[euro, expert, ttscaled=false, sfscaled=false]{mathdesign}
\usepackage{setspace}
\usepackage{csquotes}
\usepackage{listings}
\usepackage{enumerate}
\usepackage{multicol}
\usepackage{pgfplots}
\usepackage{hyperref}
\usepackage{tikz}
\usepackage{float}
\usepackage[export]{adjustbox}
\usepackage{soul}

\usepackage{dsfont}
\usepackage{bm}
\usepackage{caption}
\usepackage{subcaption}
\usepackage{multirow} 
\usepackage{makecell}

\usepackage{comment}

\usepackage{natbib}
\hypersetup{colorlinks, citecolor = red, urlcolor  = blue,}

\renewenvironment{abstract}
 {\small
  \begin{center}
  \bfseries \abstractname\vspace{-.5em}\vspace{0pt}
  \end{center}
  \list{}{%
    \setlength{\leftmargin}{10mm}
    \setlength{\rightmargin}{\leftmargin}%
  }%
  \item\relax}
 {\endlist}

\usepackage{epsfig, amssymb, multirow, algorithmic, amsmath}

\newtheorem{Remark}{Remark}[section]

\begin{document}

\Large
\begin{center}
 \textbf{Pricing and Calibration of VIX Derivatives in Mixed Bergomi Models via Quantisation}\\ \vspace{4mm} 
    \large 
Nelson Kyakutwika\footnotemark[1]\textsuperscript{*},  Mesias Alfeus\textsuperscript{2,3},  Erik Schl\"{o}gl\textsuperscript{4,5,6} 
\vspace{4mm} \\
\small 
\textsuperscript{1}Mathematics Division, Stellenbosch University,  Stellenbosch, $7600$, South Africa \\
\textsuperscript{2}Department of Statistics and Actuarial Science, Stellenbosch University,  Stellenbosch, $7600$, South Africa \\
\textsuperscript{3}National Institute for Theoretical and Computational Sciences, South Africa \\
\textsuperscript{4}School of Mathematical and Physical Sciences, University of Technology Sydney, Ultimo, NSW, $2007$, Australia \\
\textsuperscript{5}African Institute for Financial Markets and Risk Management, University of Cape Town, Rondebosch, $7701$, South Africa\\ 
\textsuperscript{6}Faculty of Science, Department of Statistics, University of Johannesburg, Auckland Park, Johannesburg, $2006$, South Africa\\  \vspace{4mm}
    \large 
June 29, 2025
\end{center}
\renewcommand{\thefootnote}{\fnsymbol{footnote}} 
\footnotetext[1]{Corresponding author. Email: \href{mailto:nelsonkyakutwika@aims.ac.za}{\textcolor{blue}{nelsonkyakutwika@aims.ac.za}}}
\renewcommand{\thefootnote}{}
\footnote{We gratefully acknowledge Bruce Bartlett for his contribution during the early stages of this work.}
\renewcommand{\thefootnote}{\arabic{footnote}} 
\normalsize
\begin{abstract}
 We apply vector quantisation within mixed one- and two-factor Bergomi models to implement a fast and efficient approach for option pricing in these models. This allows us to calibrate such models to market data of VIX futures and options. Our numerical tests confirm the efficacy of vector quantisation, making calibration feasible over daily data covering several months. This permits us to evaluate the calibration accuracy and the stability of the calibrated parameters, and we provide a comprehensive assessment of the two models. Both models show excellent performance in fitting VIX derivatives, and their parameters show satisfactory stability over time. 


\end{abstract}

\section{Introduction}
  The VIX is not only a market-implied indicator of volatility, but futures and options on the VIX are also used to hedge volatility exposure of more complex option portfolios \cite{guyon2024dispersion, bourgey2023weak}. 
 The VIX index is not tradable, but its derivatives, such as futures and options, are. In $2004$, the Chicago Board Options Exchange (CBOE) introduced VIX futures, and in $2006$, VIX options were launched for trading.   A VIX futures contract that expires at time $T_i$ is the instrument that pays $VIX_{T_i}$ at time $T_i$. The underlying asset for VIX options is the VIX futures contract, which has the same maturity as the options.   The payoffs of the VIX options are the classical ones: $(VIX_{T_i} - K)^+$ for a VIX call and $(K-VIX_{T_i})^+$ for a VIX put at time $T_i$, where $T_i$ is the VIX futures maturity and $K$ is the strike.

Mixed Bergomi models belong to the broader class of so-called Bergomi models (see \cite{bergomi2005smile}, \cite{bergomi2008smile}, \cite{bergomi2015stochastic}), which are forward variance curve models widely used by market practitioners. The term ``mixed'', also used in \cite{bourgey2023weak}, is inspired by ``mixing'' exponentials within the class of Bergomi models.  Being Markovian, Bergomi models are computationally faster compared to their rough volatility counterparts (see e.g., \cite{romer2022empirical}, \cite{jaber2022joint}), which are non-Markovian. Fast-to-compute models enable efficient pricing and calibration. In this study, we apply vector quantisation to accelerate computations in mixed Bergomi models. This makes it feasible to calibrate the models to VIX derivatives over several months of daily data, allowing us to evaluate the empirical calibration performance and parameter stability of these models.

In the literature:, the mixed two-factor Bergomi model was introduced by Bergomi in \cite{bergomi2008smile}, who observed that the mixing of two log-normal forms of the classical Bergomi model \cite{bergomi2005smile} could reproduce the upward-sloping smiles of VIX options. Since its introduction, several studies have extended this class of models. For example, Ould Aly \cite{ould2014forward} proposed a variant of the model in \cite{bergomi2008smile} that computes VIX futures and options using semi-analytic formulas and jointly calibrated it to VIX futures and puts. Further research has focused on alternative implementation techniques for mixed Bergomi models. For example, while Bergomi \cite{bergomi2008smile, bergomi2015stochastic} employs quadrature schemes to compute VIX future and option prices, Bourgey, De Marco and Gobet \cite{bourgey2023weak} derive alternative formulas for pricing in the mixed one-factor model. Similarly, Guyon \cite{guyon2022vix} derives alternative formulas for computing VIX futures within mixed Bergomi models. Despite the progress made in these studies, one aspect that has not been explored is the calibration of these models over longer data periods. Quadrature schemes, for example, as used in \cite{bergomi2008smile, bergomi2015stochastic}, are too slow for such calibration to be feasible, so a faster alternative is needed. In this study, we explore the use of vector quantisation within the class of mixed Bergomi models as a faster alternative to quadrature methods and show that it significantly improves calibration efficiency.  A recent study related to ours is  Abi Jaber, Illand, and Li \cite{jaber2022joint}, where a generic functional quantisation approach is developed for one-factor models (including a one-factor Bergomi-like model) with different kernels and applied to the SPX-VIX joint calibration problem. Our study uses quantisation to calibrate the mixed one- and two-factor Bergomi models to VIX futures and options over a longer study period. 

 Our study contributes by calibrating mixed one- and two-factor Bergomi models to VIX futures and options with maturities ranging from one week to nine months, over several months of daily data, using vector quantisation. Vector quantisation provides the computational speed necessary for such calibration, with vector quantisation being twice as fast as exact quadrature in the one-factor model and approximately $120$ times faster in the two-factor model. We evaluate the models' \textbf{static performance} by assessing their fit to the global VIX options surface and their \textbf{dynamic performance} by examining the stability of the calibrated parameters. Our empirical analysis results indicate that both models achieve superior accuracy in calibrating VIX futures and calls, with most performance errors either $0$ or close to $0$. However, the dynamic performance of the models is satisfactory but not exceptional. To the best of our knowledge, this is the first study to calibrate mixed Bergomi models over a prolonged period. Prior to calibration, we demonstrate the accuracy of vector quantisation through numerical tests for price computation. The results show relative errors smaller than $0.01\%$ for the one-factor model and less than $2\%$ for the two-factor model. 

This paper is structured as follows. Section~\ref{sec:2} describes the structure of mixed Bergomi models and sets up most of the notation. In Section~\ref{sec:3}, we derive the formulas for pricing VIX derivatives within the current model framework and test the numerical accuracy and speed of quantisation. Section~\ref{sec:calibration} calibrates the models to VIX futures and options using market data. Section~\ref{sec:parameter-stability} looks into the time evolution and stability of calibrated parameters, and Section~\ref{sec:conclusion} concludes the paper.

\section{The models: notation, definitions and structure}\label{sec:2}
\subsection{Mixed one-factor Bergomi model}
The mixed one-factor model is obtained by mixing two (log-normal) one-factor Bergomi forms with different values of the instantaneous volatility of the instantaneous forward variance. The forward variance in the resultant model is no longer log-normal but retains the Markovian property. 

Let $X = (X_t)_{t\geq 0}$ be the Ornstein-Uhlenbeck (OU) process with dynamics $$dX_t = -kX_tdt + dW_t, \quad X_0 = 0,$$
where $k \geq 0$ is the rate at which $X$ mean reverts to zero, and $W = \left( W_t \right)_{t\geq 0}$ is a one-dimensional Brownian motion defined on a risk-neutral filtered probability space $\big(\Omega, \mathcal{F}, (\mathcal{F}_t)_{t\geq 0}, \mathbb{Q}\big)$. Furthermore, for $T>t$, let $\xi_t^T$ be the instantaneous variance of SPX at time $T$ as seen from $t$. Let the forward variance in the log-normal form is given by: 
\begin{align}\label{eqn:solution-one-factor-bergomi}
\xi_t^T = \xi_0^T g^T(t, X_t),     
\end{align}
with 
\begin{align}\label{eqn:log-normal:bergomi-one}
    g^T(t, X_t) := \text{exp} \left( \omega e^{-k(T-t)}X_t - \frac{\omega^2}{2} e^{-2k(T-t)}\text{Var}(X_t)\right),
\end{align}
where $\omega \geq 0$ is the instantaneous log-normal volatility of the instantaneous variance $\xi_t^{T=t}$ and $$\text{Var}(X_t) = \frac{1-e^{-2kt}}{2k} \mathds{1}_{k>0} + t\mathds{1}_{k=0}.$$
By introducing a mixing parameter $\gamma^T \in [0,1]$ that depends on $T$, we can construct a convex combination of two exponential functions of the form in \eqref{eqn:log-normal:bergomi-one}, as demonstrated in \cite{bourgey2023weak}. The mixed form of the model, originally introduced in \cite{bergomi2008smile}, requires that the instantaneous volatility of the instantaneous variance also depends on $T$. The mixed-form analogue of \eqref{eqn:log-normal:bergomi-one} is now defined as 
\begin{align}\label{eqn:mixed-function-bergomi-one}
    f^T(t, x_t^T) =  \left(1-\gamma^T\right)\text{exp}\left(\omega^T_1x_t^T - \frac{\left(\omega^T_1\right)^2}{2}h(t,T)\right) + \gamma^T \text{exp}\left(\omega^T_2x_t^T - \frac{\left(\omega^T_2\right)^2}{2}h(t,T)\right),
\end{align}
where $$x_t^T =e^{-k(T-t)}X_t \quad \text{ and } \quad h(t, T) = e^{-2k(T-t)}\text{Var}(X_t) = \frac{e^{-2k(T-t)} - e^{-2kT}}{2k}. $$ The forward variance $\xi_t^T$ is a martingale on $[0, T]$, so its $dt$ part is zero. By It\^{o}'s lemma, the dynamics for $\xi_t^T = \xi_0^T f^T(t, x_t^T)$ read: 
\begin{align*}
 d\xi_t^T  = \xi_0^T e^{-k(T-t)}\left(\omega^T_1\left(1-\gamma^T\right) e^{\omega_1^T x_t^T - \frac{\left(\omega_1^T\right)^2}{2}h(t, T)} +  \gamma^T \omega_2^T e^{ \omega_2^T x_t^T - \frac{ \left(\omega_2^T\right)^2}{2}h(t, T)} \right) dW_t.
\end{align*}

\subsection{Mixed two-factor Bergomi model}
Unlike the case of the one-factor model, in the two-factor version, the dynamics of forward variance is driven by two OU processes: $X^1 = (X^1_t)_{t\geq 0}$ and $X^2 = (X^2_t)_{t\geq 0}$, whose dynamics are given by 
\begin{equation*}
    dX_t^l = -k_lX_t^ldt + dW_t^l, \quad X_0^l = 0, \quad l\in\{1, 2\},
\end{equation*}
where, without loss of generality, we set  $k_1 > k_2$.  Let $\rho$ be the correlation between $W^1$ and $W^2$ such that the correlation between $X^1$ and $X^2$ is $v_t^{1,2}/\sqrt{v_t^1 v_t^2}$, where $$v_t^1 =  \frac{1-e^{-2k_1t}}{2k_1}, \quad v_t^2 = \frac{1-e^{-2k_2t}}{2k_2}, \text{ and} \quad  v_t^{1,2} = \rho \frac{1-e^{-(k_1+k_2)t}}{k_1 + k_2}.$$
Following \cite{bergomi2015stochastic}, we define a Markov process $$\lambda_t^T = \alpha_\theta \left[(1-\theta) e^{-k_1(T-t)}X_t^1 + \theta e^{-k_2(T-t)}X_t^2\right],$$
where $\theta \in [0,1]$ is called a mixing parameter and $\alpha_\theta = 1/ \sqrt{(1-\theta)^2 + \theta^2 + 2\rho \theta (1-\theta)}$ is a normalising factor which ensures that $\omega$ is the instantaneous volatility  of $\xi_t^t$. Then, the (log-normal) forward variance in the classical two-factor Bergomi model is given by $$\xi_t^T = \xi_0^Tg^T\left(t, X_t^1, X_t^2\right),$$
where 
\begin{align}\label{eqn:log-normal:bergomi-two}
    g^T(t, X_t^1, X_t^2): = \text{exp}\left( \omega \lambda_t^T - \frac{\omega^2}{2}\chi(t, T)\right),
\end{align}
with 
\begin{align*}
\chi(t, T)& = \int_{T-t}^{T} \alpha_\theta^2 \left[ (1-\theta)^2 e^{-2k_1\tau} + \theta^2 e^{-2k_2\tau}  + 2\rho\theta(1-\theta)e^{-(k_1 + k_2)\tau}\right]d\tau \\
& = \alpha_\theta^2 \left( (1-\theta)^2e^{-2k_1(T-t)}v_t^1  + \theta^2 e^{-2k_2(T-t)} v_t^2 + 2\theta(1 -\theta) e^{-(k_1 + k_2)(T-t)} v_t^{1,2} \right).
\end{align*}
Using the mixing parameter $\gamma^T \in [0, 1]$ as for the one-factor model, the convex combination of two exponential functions of the form  \eqref{eqn:log-normal:bergomi-two} is defined as 
\begin{align}\label{eqn:mixed-function-bergomi-two}
    f^T\left(t, \lambda_t^T\right) =  (1-\gamma^T)\text{exp}\left(\omega^T_1\lambda_t^T - \frac{\left(\omega^T_1\right)^2}{2}\chi(t,T)\right) + \gamma^T \text{exp}\left(\omega^T_2\lambda_t^T - \frac{\left(\omega^T_2\right)^2}{2}\chi(t,T)\right),
\end{align}
and, by the multivariate extension of It\^{o}'s lemma, the dynamics  for $\xi_t^T = \xi_0^T f^T(t, \lambda_t^T)$ are given by
\begin{align*}
 d\xi_t^T & 
= \xi_0^T \alpha_\theta \left(\omega_1^T\left(1-\gamma^T\right) e^{\omega_1^T \lambda_t^T - \frac{\left(\omega_1^T\right)^2}{2}\chi(t, T)} + \gamma^T\omega_2^T e^{ \omega_2^T \lambda_t^T - \frac{ \left(\omega_2^T\right)^2}{2}\chi(t, T)} \right) \times \\
& \quad \left((1-\theta)e^{-k_1(T-t)} dW_t^1 + \theta e^{-k_2(T-t)} dW_t^2 \right).
\end{align*}

We adjust the parametrisation of the mixed two-factor model from the conventional form, introduced in \cite{bergomi2008smile} and used in subsequent studies such as \cite{bergomi2015stochastic}. In the conventional parametrisation, two parameter sets are used: $(k_1, k_2, \theta, \rho, v = \omega/2)$ and $(\gamma_T, \beta_T, \zeta_T)$, where $\beta_T$ and $\zeta_T$ are volatility-of-volatility (vol-of-vol) smile parameters. Each parameter set is calibrated separately, with the first set calibrated in the classical two-factor Bergomi model and the second set calibrated in the mixed two-factor Bergomi model for given values of the first parameter set. The first set is calibrated to establish a direct handle on the term structure of the vol-of-vol, while the second set is calibrated on the vanilla smile. Inspired by the approach of \cite{bourgey2023weak} with the one-factor model, we propose a new parametrisation that enables calibration of the parameters of the mixed two-factor Bergomi model in a single step and directly on the vanilla smile. In our new set-up~\eqref{eqn:mixed-function-bergomi-two}, we use the parameter sets $(k_1, k_2, \theta, \rho)$ and $(\gamma^T, \omega_1^T, \omega_2^T)$, with the parameter $v$ from the conventional parametrisation absorbed by $\omega_1^T$ and $\omega_2^T$. Our adjustment allows us to set the parameter set $(k_1, k_2, \theta, \rho)$ a priori--for example, as in \cite{ould2014forward}--and calibrate the parameter set $(\gamma^T, \omega_1^T, \omega_2^T)$ using liquid instruments, namely VIX futures and options.

\section{Pricing VIX derivatives}\label{sec:3}
\subsection{The VIX index}
The VIX at time $t\geq 0$ is the implied volatility of a $30$-day log-contract on the SPX index starting at time $t$, computed by the CBOE by replication using market prices of listed S\&P $500$ options: 
\begin{align}
    \text{VIX}_t^2: & = -\frac{2}{\Delta}\text{Price}_t\left[ \log\left(\frac{S_{t+\Delta}}{F_t^{t+\Delta}}\right)\Big|\mathcal{F}_t \right] \label{eqn:defn-vix} \\
    & = \frac{2e^{r\Delta}}{\Delta}\left( \int_0^{F_t^{t+\Delta}}\frac{P(t, t+\Delta, K)}{K^2}dK + \int_{F_t^{t+\Delta}}^\infty \frac{C(t, t+\Delta, K)}{K^2} dK\right), \label{eqn:defn-vix-cboe}
\end{align}
where $\Delta = 30$ calender days, $P(t, t+\Delta, K)$ (respectively $C(t, t+\Delta, K)$) is the discounted market price at time $t$ of a put (respectively call) option of maturity $t+\Delta$ and strike $K$ on the S\&P $500$ index, $F_t^{t+\Delta}$ is the forward price of the S\&P $500$ index for maturity $t+\Delta$ observed at $t$, and $r$ is the risk-free interest rate. 

In a model-free manner, the CBOE replicates \eqref{eqn:defn-vix} using \eqref{eqn:defn-vix-cboe}, which is derived from the Carr-Madan formula, as studied in \cite{carr1998towards}. In continuous-time stochastic volatility models, 
\begin{align}\label{eqn:VIX-equations}
      \text{VIX}_t^2: & = -\frac{2}{\Delta}\mathbb{E}\left[ \log\left(\frac{S_{t+\Delta}}{F_t^{t+\Delta}}\right)\Big|\mathcal{F}_t \right]  = \mathbb{E}\left[\frac{1}{\Delta}\int_t^{t+\Delta} \xi^T_T dT \Big|\mathcal{F}_t\right] = \frac{1}{\Delta}\int_t^{t+\Delta}\xi_t^TdT.
\end{align}

Following \cite{romer2022empirical, jaber2022quintic, cuchiero2023joint}, in our numerical experiments, we work with the convention where \eqref{eqn:VIX-equations} is scaled with the factor $100^2$, that is,
\begin{align}\label{eqn:VIX-equation-scaling-factor}
     VIX_t^2 = \frac{100^2}{\Delta}\int_t^{t+\Delta}\xi_t^TdT.
\end{align}
The scaling factor ensures that $VIX_t$ is expressed as a percentage. Importantly, the results of this study remain valid with or without the scaling factor.  

\subsection{Overview of vector quantisation} \label{sec:3.2}
Here, we give an overview of quantising a random variable $Y\in \mathbb{R}^d$ (vector quantisation). For details, see \cite{pages2003optimal, pages2015recursive, jaber2022joint, callegaro2017pricing, callegaro2014pricing} and references therein. Let $Y$ be defined on the probability space $(\Omega, \mathcal{F}, \mathbb{P})$, with finite $r$th moment. Quantising $Y$ means approximating the continuum of values of $Y$ by a discrete random variable that takes values in a set of cardinality $N$. Quantising \(Y\) corresponds to projecting it onto the grid $\Gamma = \{ y_1, \ldots, y_N\} \subset \mathbb{R}^d$ using the nearest neighbourhood projection. The Borel function $q: \mathbb{R}^d \longrightarrow \Gamma$, projecting $Y$ onto $\Gamma$ is called the quantisation function or $N$-quantiser.   The notations $\widehat{Y}^\Gamma$ or $\widehat{Y}$ are usually used as alternatives to the function $q(Y)$. Our goal is to look for the $L^2$-optimal $N$-quantiser $\widehat{Y}^{\Gamma^*}$ that minimises the $L^2$-mean quantisation  error which is given by             $$e_N(Y, \Gamma) : = \|Y-\widehat{Y}^\Gamma \|_2 = \| \underset{1 \leq j \leq N}{\text{min}} |Y -y_j|\ \|_2, $$ where $\| Y\|_2\colon  = \big[ \mathbb{E}(|Y|^2) \big]^{1/2}.$ The mean quantisation  error can be generalised to the $L^r$-norm, for $r > 0$, but the computed values used in our study utilise the $L^2$-norm (quadratic case). 
 The optimal quantiser $\widehat{Y}^{\Gamma^*}$ corresponds to the optimal Borel partition of $\mathbb{R}^d$ so that the optimal quantiser is defined as  $$\widehat{Y}^{\Gamma^*} = \sum_{j = 1}^N y_j \mathds{1}_{C_j(\Gamma^*)}(Y), $$ where the (optimal) Borel partition $\{C_j(\Gamma^*) \}_{j = 1, \ldots, N}$ is called the Vorono\"{i} partition induced by $\Gamma^*$.
Considering the ordered quantisers $y_1 < y_2 < \ldots < y_{N-1} < y_N$, the Vorono\"{i} cells, for  $j = 1, \ldots, N$, are defined by  $$C_j(\Gamma^*) = \big[y_{j -\frac{1}{2}}, y_{j + \frac{1}{2}}\big), \quad y_{j \pm \frac{1}{2}} = \frac{y_j + y_{j\pm 1}}{2}, \quad y_{\frac{1}{2}} = \text{inf}\big(\text{supp}(\mathbb{P}_Y) \big), \quad y_{N+\frac{1}{2}} = \text{sup}\big( \text{supp}(\mathbb{P}_Y)\big),$$ where $\text{supp}(\mathbb{P}_Y)$ denotes the support of the probability distribution of $Y$, $\mathbb{P}_Y$.   The quantiser $\widehat{Y}^{\Gamma^*}$ corresponds to the probabilities $\{p_1,\ldots,p_N\}$ where $$p_j = \mathbb{P}(\widehat{Y}^{\Gamma^*} = y_j) = \mathbb{P}(Y \in C_j(\Gamma^*)).$$ As a result, integrals of the form $\mathbb{E}[f(Y)]$ are approximated by the finite sum
$$\mathbb{E}[f(Y)] \approx  \mathbb{E}[f(\widehat{Y}^{\Gamma^*})] = \sum_{j= 1}^Nf(y_j)p_j.$$ Quadratic optimal $N$-point quantisers for standard Gaussian variables have been computed offline and are available at the Quantisation  Website \cite{quantization}. In our study, we use vector quantisation specifically but occasionally refer to it simply as quantisation.

\subsection{Pricing VIX derivatives via quantisation}
\subsubsection{Pricing VIX futures}
In the case of the one-factor model, the random variable to be quantised follows the OU process $X$. To quantise $X_t$, a one-dimensional optimal $N$-point quantiser, $z_N$, of the standard normal distribution is scaled by the standard deviation of $X_t$: \[X_t = \sqrt{\frac{1-e^{-2kt}}{2k}}\cdot z_N.\] Once the $X_t$ has been quantised, the function $f^T(t, x_t^T)$ can be computed for a given maturity $T_i$ and known parameters, and,  subsequently, quantised values of  $VIX^2_{T_i}$ can be obtained from 
\begin{align}\label{eqn:remark1-quantisation }
    VIX_{T_i}^2 = \frac{1}{\Delta}\int_{T_i}^{T_i+\Delta}\xi_{T_i}^TdT.
\end{align}
By denoting $V_j^i$, for $j = 1, \ldots, N$,  as the quantised values for $VIX_{T_i}^2$, the approximation $$\mathbb{E}[VIX_{T_i}] \approx \mathbb{E}\left[\sqrt{V_j^i}\right]$$ is used to compute VIX futures. Thus, the VIX futures of  expiry $T_i$ (observed at $t=0$) is obtained from
\begin{align}\label{eqn:5}
   F_0^{T_i} =  \mathbb{E}[ VIX _{T_i}] \approx \sum_{j=1}^N\sqrt{V^i_j}p_j.
\end{align}
In the two-factor model, we obtain $V_j^i$  by quantising the bivariate normal distribution: $$ \bm{X}_t: =  \begin{pmatrix}
X_t^1 \\
X_t^2
\end{pmatrix} \sim \mathcal{N}(\bm{0}, \bm{\Sigma}_t), \quad \bm{\Sigma}_t = \begin{pmatrix}
v_t^1 & v_t^{1,2} \\
v_t^{1,2} & v_t^2
\end{pmatrix},
$$ with the help of the Cholesky decomposition. See the remark below. 
\begin{Remark}[Cholesky decomposition]\label{remark Cholesky}
Let $Z_1$ and $Z_2$ be two independent standard normal random variables, and define 
$$Y^1 = Z_1 \quad \text{ and } \quad Y^2 =  \rho_{12} Z_1 + \sqrt{1-\rho_{12}^2}\cdot Z_2$$ where $\rho_{12} \in [-1, 1]$. Then,   $Y^1$ and $Y^2$ are two standard bivariate normal random variables with  $\text{Corr}(Y^1, Y^2) = \rho_{12}$. 
\end{Remark}
Consequently, the quantisation of the two random variables following correlated OU processes proceeds as follows. Let $z_N^1$ and $z_N^2$ be the N-point quantisers of the standard bivariate normal distribution, and define \[y_N^1 = z^1_N, \quad y_N^2 = \rho_{12}z_N^1 + \sqrt{1-\rho_{12}^2}\cdot z_N^2.\]  Then, $X_t^1$ and $X_t^2$ are given by \[X_t^1 = \sqrt{\frac{1-e^{-2k_1t}}{2k_1}} \cdot y_N^1, \quad X_t^2 = \sqrt{\frac{1-e^{-2k_2t}}{2k_2}} \cdot y_N^2,\] so that \( \text{Corr}(y^1_N, y_N^2) = \text{Corr}(X_t^1, X_t^2) = \rho_{12} \).

\subsubsection{Pricing VIX calls and puts}
At time $t=0$, we obtain the  prices of VIX calls and puts for maturity $T_i$ using the classical formulas:
\begin{align*}
C_0^{T_i} = e^{-rT_i}\mathbb{E}[(VIX_{T_i} - K)^+] \approx  e^{-rT_i}\sum_{j=1}^{N} \left(\sqrt{V^i_j} - K\right)^+p_j,     
\end{align*}
\begin{align*}
    P_0^{T_i} = e^{-rT_i} \mathbb{E}[(K- VIX_{T_i})^+] \approx e^{-rT_i} \sum_{j=1}^{N} \left(K- \sqrt{V^i_j}\right)^+p_j,
\end{align*}
where $r$ is the risk-free interest rate and $K$ is a VIX call/put strike.

\subsection{Accuracy and speed of quantisation: numerical tests}\label{subsec: accuracy-quantisation}
In this section, we assess the accuracy of quantisation in computing VIX futures and option prices within our models. Reference prices are obtained using a two-dimensional quadrature--one dimension in time and one in space--for the one-factor model, and a three-dimensional quadrature--one in time and two in space--for the two-factor model. Our findings demonstrate that quantisation is highly accurate, with results for the one-factor model being graphically indistinguishable from those produced by exact quadrature. 

To compute VIX futures and option prices via quantisation in the one-factor model, we use the $1000$-quantiser of the standard univariate normal distribution available at \cite{quantization}. For the two-factor model, we employ the $1450$-quantiser of the standard bivariate normal distribution, also available at \cite{quantization}. We observed that $N=1000$ in the one-factor model achieves a good balance between accuracy and computational cost. For the two-factor model, we increased $N$ to its maximum, as the website offers the $1450$-quantiser as the largest available quantiser for the standard bivariate normal distribution. 

We now describe the computation of reference prices via quadrature. For the one-factor model, note that $VIX^2$ is a function of $X$. Consequently, pricing VIX derivatives with a payoff function $\Phi$ can be accomplished by integrating with respect to the standard univariate Gaussian density. Since \( VIX^2_{T_i} = h(X_{T_i}) \),
\begin{align*}
    \mathbb{E}[\Phi(VIX_{T_i})] = \mathbb{E}\left[ \Phi\left( \sqrt{h(X_{T_i})} \right) \right]  = \frac{1}{\sqrt{2\pi}}\int_\mathbb{R}\Phi\left( \sqrt{h(\sigma_{X_{T_i}} z)} \right)e^{-z^2/2} \, dz,
\end{align*}
where \( \sigma_{X_{T_i}} \) is the standard deviation of \( X_{T_i} \) and \( z \sim \mathcal{N}[0, 1] \). We perform the time integral in \( h(\sigma_{X_{T_i}}z) \) and the integral with respect to the standard Gaussian (space integral) using MATLAB's \texttt{integral} function, with MATLAB’s default absolute and relative error tolerances.  

In the two-factor model, where \( VIX_{T_i}^2 = h(X^1_{T_i}, X^2_{T_i}) \),
\begin{align*}
  \mathbb{E}[\Phi(VIX_{T_i})] & = \mathbb{E}\left[\Phi\left(\sqrt{h(X^1_{T_i}, X^2_{T_i})} \right) \right] \\
                 & =   \frac{1}{2\pi \sqrt{1-\rho_{12}^2}}\int_{\mathbb{R}^2} \Phi\left( \sqrt{h(\sigma_{X^1_{T_i}}z_1, \sigma_{X^2_{T_i}}z_2)} \right) \, \times \\ 
                 & \quad \exp \left( -\frac{1}{2(1-\rho^2_{12})}(z_1^2 + z_2^2 - 2\rho_{12}z_1z_2) \right) \, dz_1 \, dz_2,
\end{align*}
where \( \sigma_{X^1_{T_i}} \) and \( \sigma_{X^2_{T_i}} \) are the standard deviations of \( X^1_{T_i} \) and \( X^2_{T_i} \), respectively, and
\begin{align*}
  \begin{pmatrix}
    z_1 \\
    z_2
  \end{pmatrix} \sim \mathcal{N}\left[   \begin{pmatrix}
    0 \\
    0
  \end{pmatrix},   \begin{pmatrix}
    1 & \rho_{12} \\
    \rho_{12} & 1
  \end{pmatrix}  \right].
\end{align*}
In \( h(\sigma_{X^1_{T_i}}z_1, \sigma_{X^2_{T_i}}z_2) \), we evaluate the single time integral using \texttt{integral} and integrate with respect to the standard bivariate normal distribution using \texttt{integral2}, with MATLAB’s default error tolerances. 

For pricing VIX futures, we use \( \Phi(U) = U \), and for calls and puts, we, respectively, use \( \Phi(U) = (U-K)^+ \) and \( \Phi(U) = (K-U)^+ \), where \( K \) is the strike. All integrals are truncated to the interval \( [-10, 10] \). Alternative quadrature schemes, such as those in \cite{bergomi2015stochastic, bourgey2023weak}, may also be used, where a combination of Gauss-Legendre and Gauss-Hermite is employed. 

\begin{figure}[h!]
    \centering
    \begin{minipage}{0.48\textwidth}
        \centering
        \includegraphics[width=\textwidth]{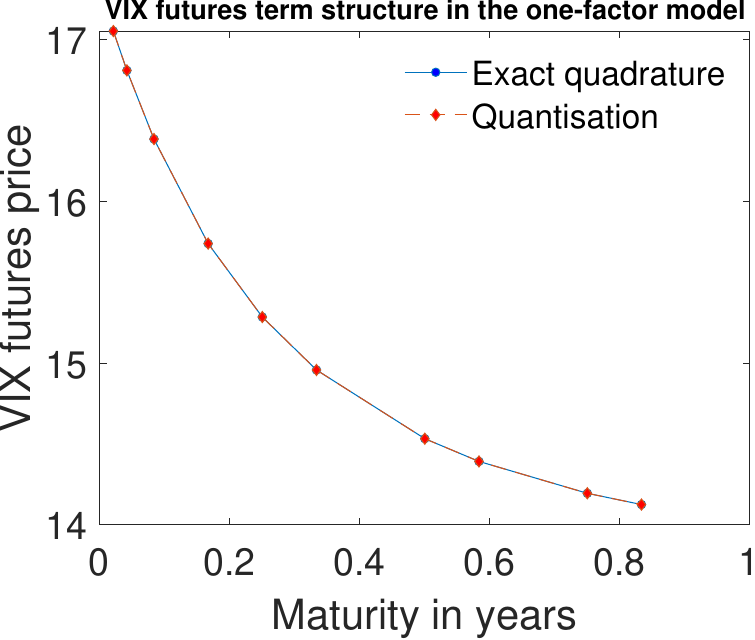}
    \end{minipage} \hspace{0.3cm}
    \begin{minipage}{0.48\textwidth}
        \centering
        \includegraphics[width=\textwidth]{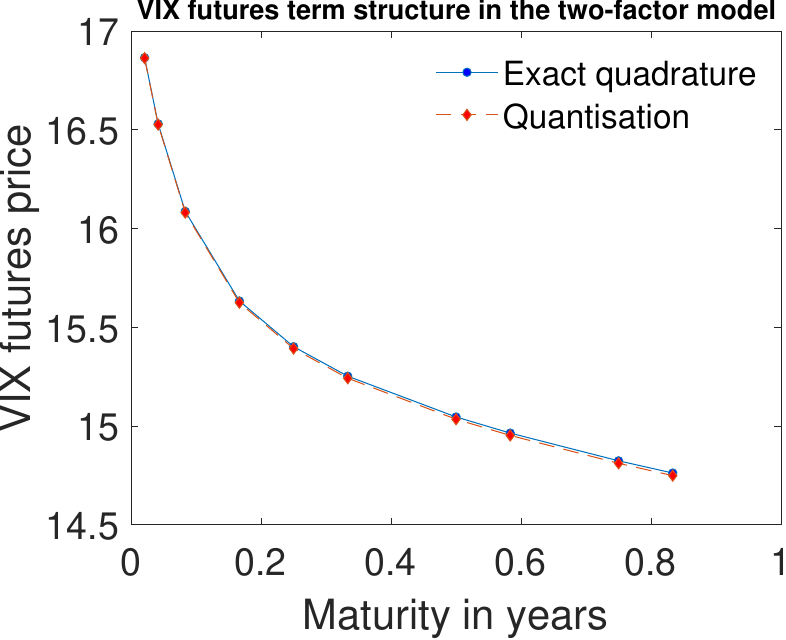}
    \end{minipage}
    \\ \vspace{0.4cm}
    \begin{minipage}{0.48\textwidth}
        \centering
        \includegraphics[width=\textwidth]{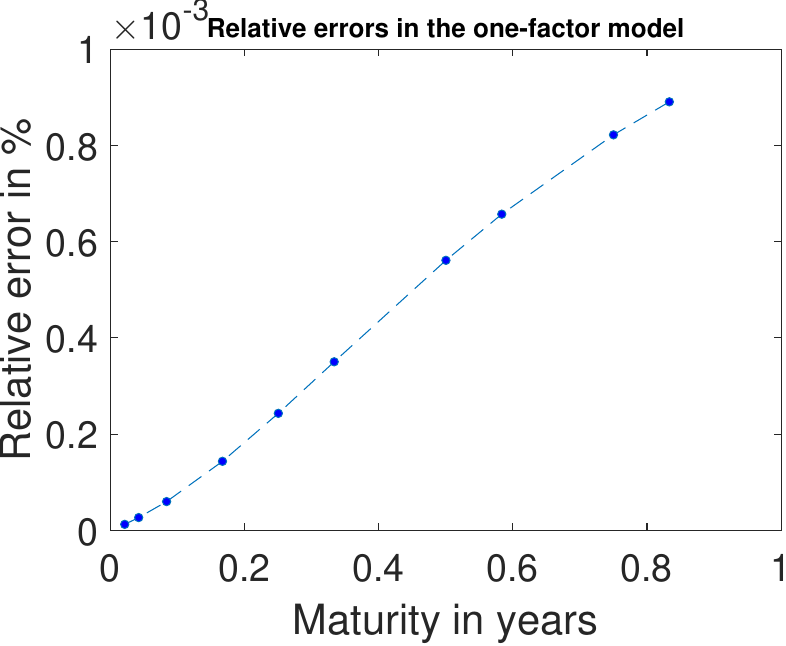}
    \end{minipage} \hspace{0.3cm}
    \begin{minipage}{0.48\textwidth}
        \centering
        \includegraphics[width=\textwidth]{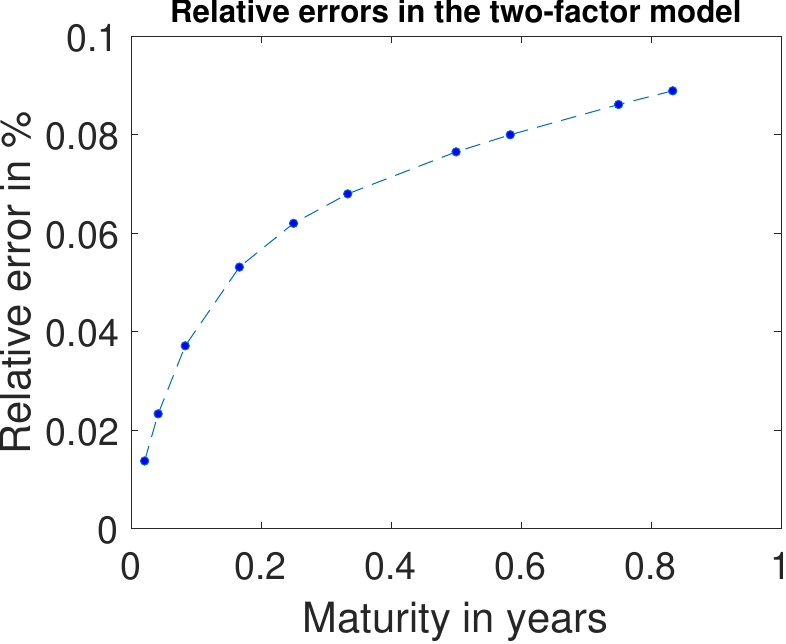}
    \end{minipage}
    \caption{\textit{Top:} VIX futures term structure in mixed Bergomi models, exact quadrature (blue) versus quantisation  (red). \textit{Bottom:} Relative errors between exact quadrature and quantisation. }
     \label{fig:term-structure-futures} 
\end{figure}

\begin{figure}[h!]
    \centering
    \begin{minipage}{0.48\textwidth}
        \centering
        \includegraphics[width=\textwidth]{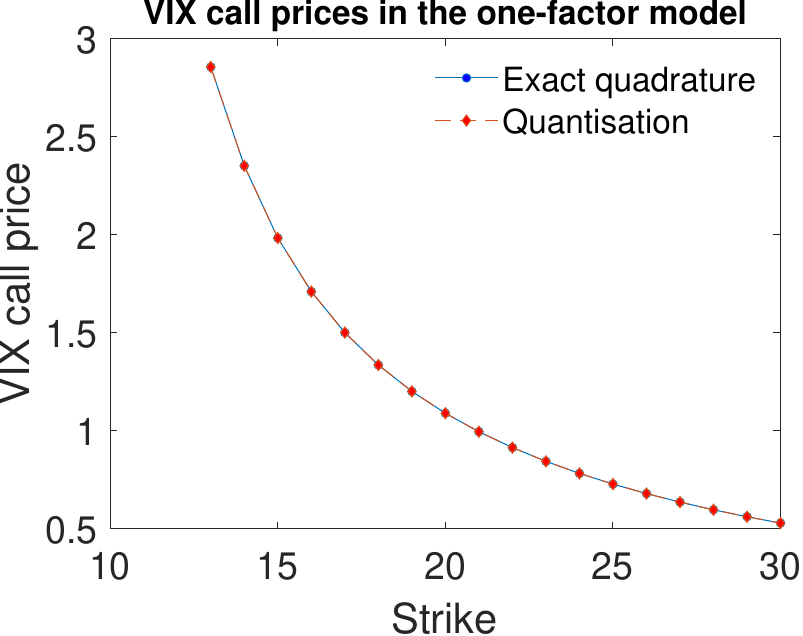}
    \end{minipage} \hspace{0.3cm}
    \begin{minipage}{0.48\textwidth}
        \centering
        \includegraphics[width=\textwidth]{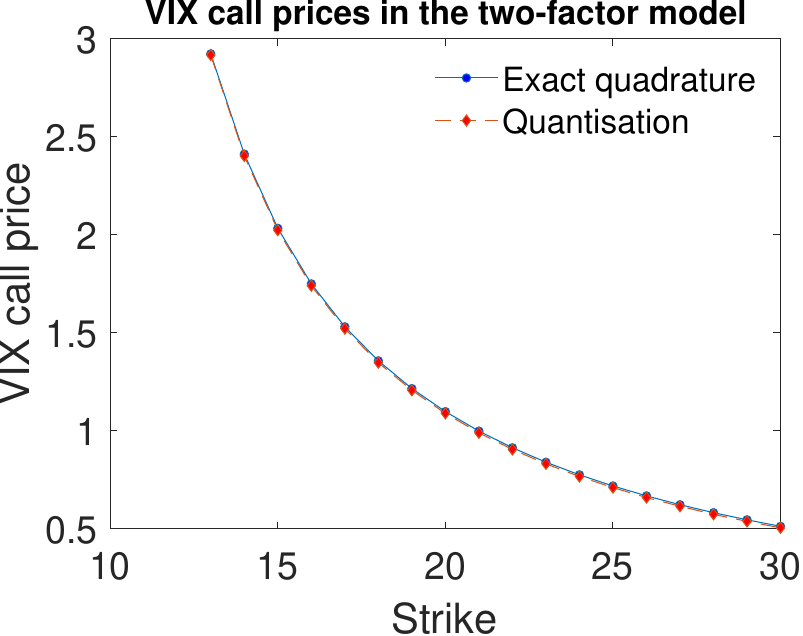}
    \end{minipage}
    \\ \vspace{0.4cm}
    \begin{minipage}{0.48\textwidth}
        \centering
        \includegraphics[width=\textwidth]{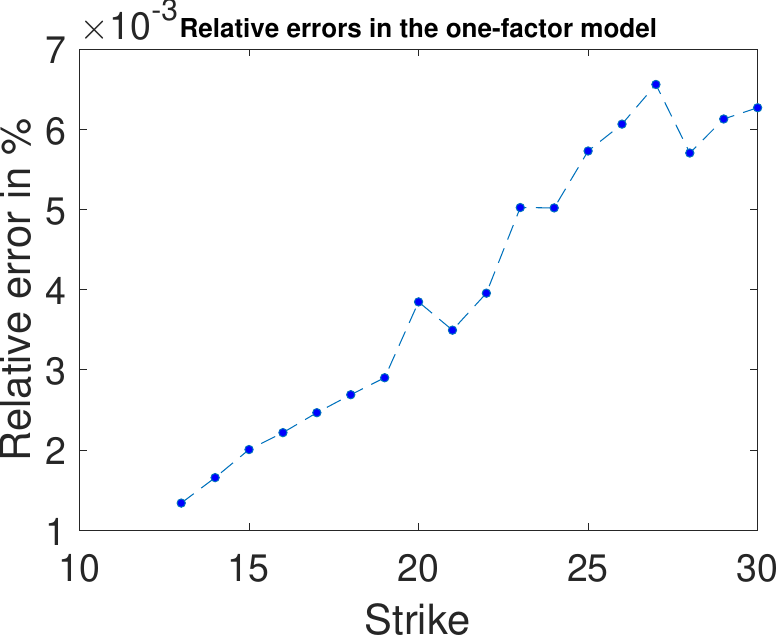}
    \end{minipage} \hspace{0.3cm}
    \begin{minipage}{0.48\textwidth}
        \centering
        \includegraphics[width=\textwidth]{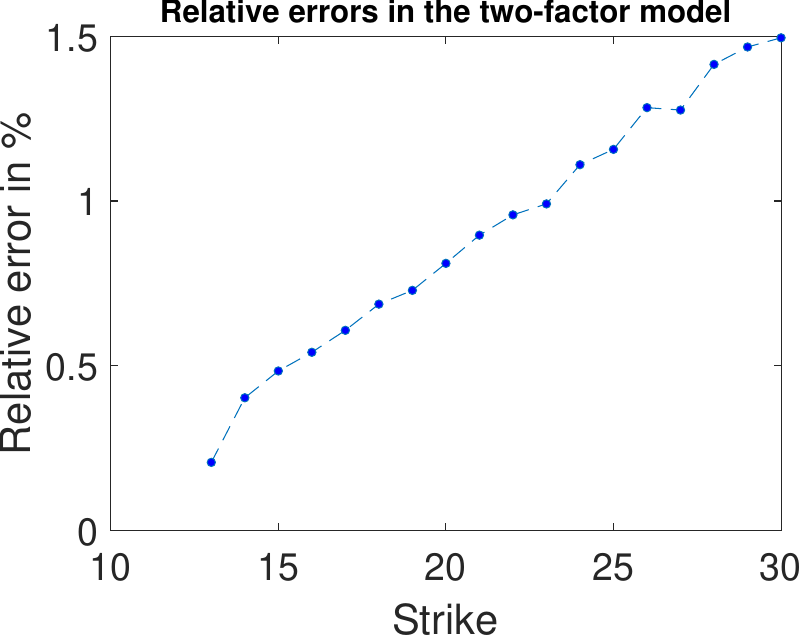}
    \end{minipage}
    \caption{\textit{Top:} VIX call prices in mixed Bergomi models, exact quadrature (blue) versus quantisation  (red). \textit{Bottom:} Relative errors between exact quadrature and quantisation.}
    \label{fig:calls-pricing} 
\end{figure}

\begin{figure}[h!]
    \centering
    \begin{minipage}{0.48\textwidth}
        \centering
        \includegraphics[width=\textwidth]{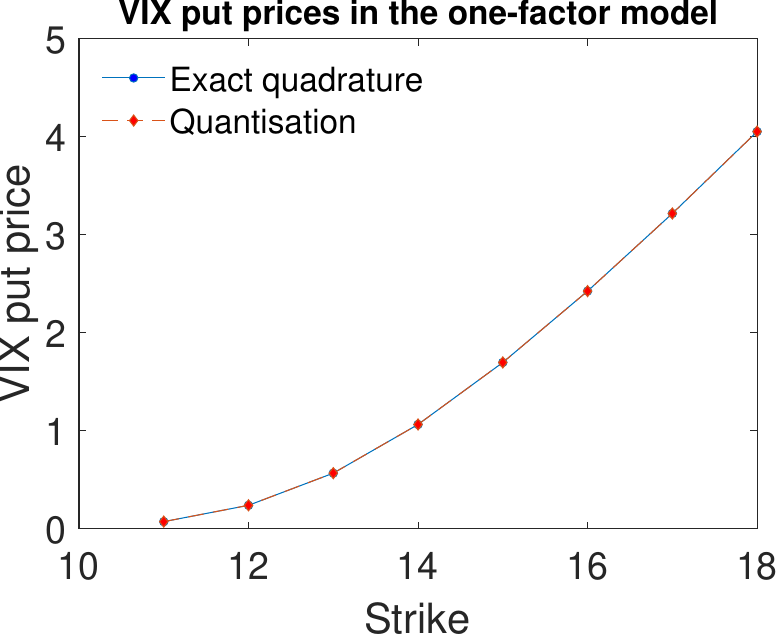}
    \end{minipage} \hspace{0.3cm}
    \begin{minipage}{0.48\textwidth}
        \centering
        \includegraphics[width=\textwidth]{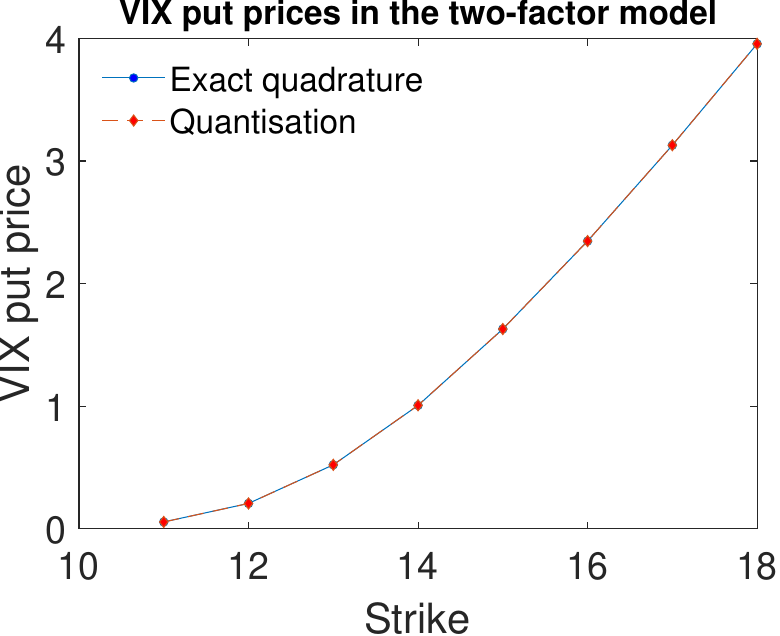}
    \end{minipage}
    \\ \vspace{0.4cm}
    \begin{minipage}{0.48\textwidth}
        \centering
        \includegraphics[width=\textwidth]{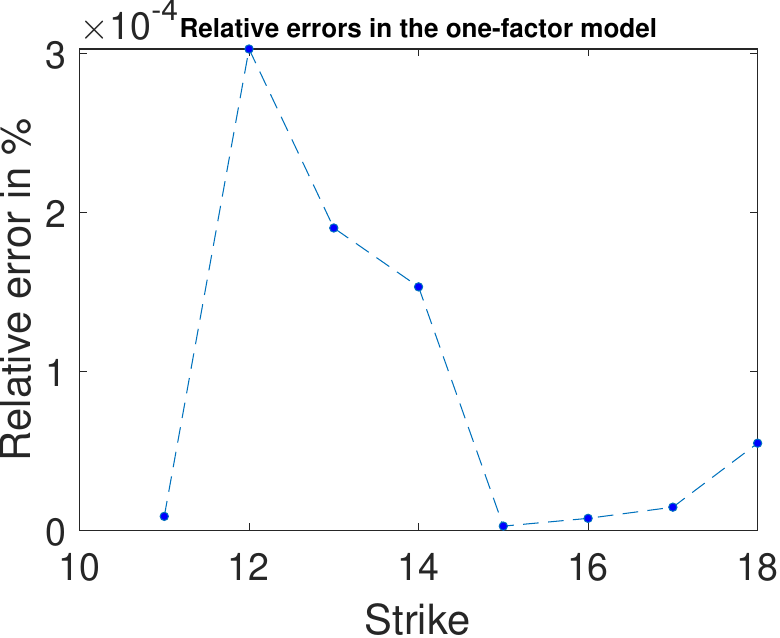}
    \end{minipage} \hspace{0.3cm}
    \begin{minipage}{0.48\textwidth}
        \centering
        \includegraphics[width=\textwidth]{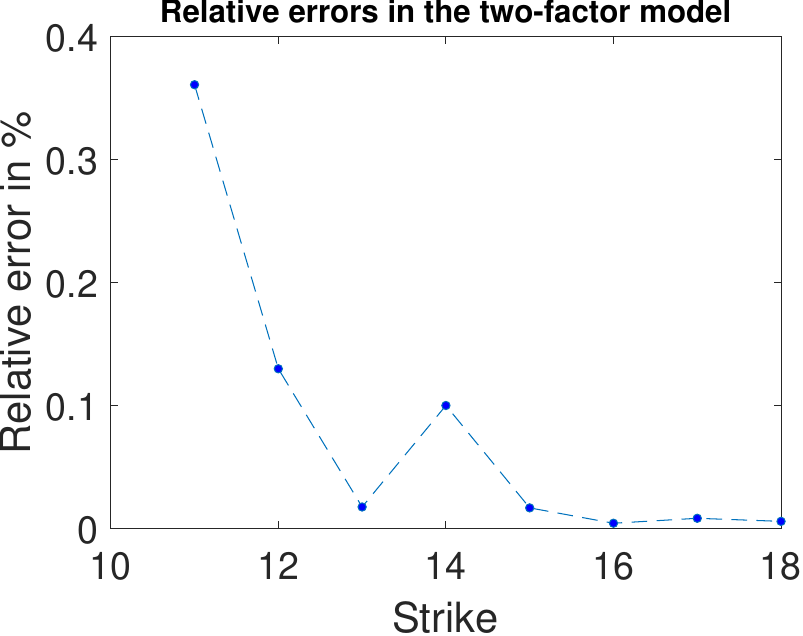}
    \end{minipage}
    \caption{\textit{Top:} VIX put prices in mixed Bergomi models, exact quadrature (blue) versus quantisation  (red). \textit{Bottom:} Relative errors between exact quadrature and quantisation.}
    \label{fig:puts-pricing} 
\end{figure}

In both the quantisation and exact quadrature approaches, for both models, we compute (a) a term structure of $10$ VIX futures (Figure~\ref{fig:term-structure-futures}), (b) $18$ call option prices (Figure~\ref{fig:calls-pricing}), and (c) $8$ put option prices (Figure~\ref{fig:puts-pricing}). For the futures, maturities range from $1$ week to $10$ months, while for the options, a maturity of $3$ months is used. We assume a flat initial term structure of forward instantaneous variances, $\xi_0^T$, fixed at $0.03$ in all cases. In the one-factor model, the at-the-money strike price for the options is computed in the model as $15.29$, whereas in the two-factor model, it is  $15.40$. The range of moneyness for the calls matches that used in the calibration in Section~\ref{sec:calibration}, [$90\%, 200\%$]. Realistic parameters, obtained from calibrating one of the slices in Section~\ref{sec:calibration}, are used: for the one-factor model, $\gamma^T = 0.61$, $\omega_1^T = 5.53$, $\omega_2^T = 0.69$, and $k = 1$; for the two-factor model, $k_1 = 7.54$, $k_2 = 0.24$, $\rho = 0.7$, $\theta = 0.23$, $\gamma^T = 0.60$, $\omega_1^T = 9.12$, and $\omega_2^T = 1.10$. These parameters are used in all the Figures~\ref{fig:term-structure-futures}, \ref{fig:calls-pricing}, and \ref{fig:puts-pricing}.

 As shown in Figures~\ref{fig:term-structure-futures}, \ref{fig:calls-pricing}, and \ref{fig:puts-pricing}, quantisation achieves high accuracy levels. For the one-factor model, the relative errors are below $0.001\%$, $0.007\%$, and $0.00035\%$ for the term structure of VIX futures, VIX calls, and VIX puts, respectively. In fact, quantisation prices from the one-factor model are so accurate that they are visually indistinguishable from those obtained through exact quadrature. For the two-factor model, the relative errors are below $0.1\%$, $1.5\%$, and $0.4\%$ for the term structure of VIX futures, VIX calls, and VIX puts, respectively. Notably, the term structure of VIX futures is decreasing due to the flat initial term structure of forward variances. An increasing term structure of VIX futures could be achieved by using an increasing initial term structure of forward variances.

In terms of computational efficiency, quantisation outperforms exact quadrature. In the one-factor model, quantisation is twice as fast as exact quadrature when computing the VIX future and call prices in Figure~\ref{fig:calls-pricing}.
For the same computations in Figure~\ref{fig:calls-pricing}, quantisation  is approximately $120$ times faster than exact quadrature in the two-factor model.

Notably, in the one-factor model, quantisation achieves higher accuracy relative to exact quadrature than the approximate price expansion formulas derived in \cite{bourgey2023weak}. Our two-factor model approximates VIX futures and calls slightly from below, particularly for longer maturities and higher strikes. Approximations of VIX futures or calls from below while using quantisation have been noted in the literature; for instance, Bonesini and Jacquier \cite{bonesini2021functional} report this behaviour when using functional quantisation to price VIX futures and options in the rough Bergomi model. However, our computed VIX future and option prices exhibit greater accuracy than those reported in \cite{bonesini2021functional}. In \cite{callegaro2017pricing}, the first study to employ recursive marginal quantisation for pricing under stochastic volatility, the results indicate that quantisation approximates benchmark prices of vanilla calls and implied volatilities slightly from below in several models; however, this study does not address VIX futures and options.

\section{Joint calibration of VIX futures and options}\label{sec:calibration}

\subsection{Dataset}
The dataset consists of the daily bid and ask quotes on VIX call and put options over $105$ trading days, from \(2\) January \(2024\) to \(31 \) May \(2024\). 
We apply some of the standard exclusion filters: we remove options that either (1) have a bid price of zero or (2) mature in fewer than $7$ days. We calculate mid options prices as the average of the regular trading hours' end-of-day bid and ask quotes, and use them to obtain the interest rate ($r$) and VIX future price ($F$) for each slice. Using put-call parity for options on futures contracts, we determine $r$ and $F$ by finding their values that minimise the sum of the squared differences  $$\sum_{i} \left(C_i - P_i - e^{-rT}(F - K_i)\right)^2,$$ done separately for each maturity. We calibrate the models over a moneyness (call option's strike divided by the VIX futures price) range of $[90\%, 200\%]$.

\subsection{Calibration}
We carried out a joint calibration of VIX futures and call options for  $105$ daily VIX options surfaces with a total of $1213$ maturity slices using the quantisation approach tested in Section~\ref{subsec: accuracy-quantisation}. Since our models consist of maturity-dependent parameters, calibration proceeds slice by slice. We jointly calibrated the models to the VIX futures and VIX calls for every slice. To speed up computations, we used MATLAB's \texttt{parfor} function that enables the distribution of daily VIX options surfaces among the cores of the computer for the calibration of the daily surfaces to proceed in parallel. Additionally, instead of using \texttt{integral} like in Section~\ref{subsec: accuracy-quantisation}, we used Gauss-Legendre quadrature to compute the one-dimensional time integral available in both models. Gauss-Legendre is accessible via MATLAB's function \texttt{lgwt} (see \cite{von2004legendre}). Using  \texttt{lgwt} with $20$ nodes is accurate enough and about $10$ times faster than the use of the \texttt{integral} function, measured by computing the future and call prices in Figure~\ref{fig:calls-pricing} using the two-factor model. 

 Calibration proceeds as follows. For each slice with maturity $T_i$, we seek to determine the parameters $(\gamma^T, \omega_1^T, \omega_2^T)$ as well as the initial forward variance $\xi_0^T$, for $T \in [T_i, T_i + \Delta]$,  such that (a) the market price of 
the VIX futures expiring at $T_i$ is matched as closely as possible and (b) call options maturing at $T_i$ stay within the bid-ask corridor as much as possible. That is, $$\mathbb{E}[VIX_{T_i}] = F_\text{mkt}(T_i) \quad \text{and} \quad C_\text{mkt}^\text{bid}(T_i, K_j) \leq  e^{-r_iT_i}\mathbb{E}[(VIX_{T_i} - K_j)^+] \leq C_\text{mkt}^\text{ask}(T_i, K_j),$$ where $F_{\text{mkt}}(T_i)$  denotes the market price of the VIX futures expiring at $T_i$ and  $C_\text{mkt}^\text{ask/bid}(T_i, K_j)$  represents market call prices for  maturity $T_i$ and strike $K_j$.

Let $m$ denote the number of call prices for maturity $T_i$, and let $\Theta$ represent the admissible set of model parameters. Then, the optimisation problem to be solved is 
\begin{gather}\label{eqn:objective-function}
     \underset{\gamma^{T}, \omega_1^T, \omega_2^T, \xi_0^T \in \Theta}{\text{argmin}} \Bigg \{\left( \frac{F_{\Theta}(T_i) - F_{\text{mkt}}(T_i)}{F_\text{mkt}(T_i)}  \right)^2  + \nonumber  \\ \frac{1}{m}\sum_{j = 1}^m \left( \text{max} \left\{\frac{C_\Theta(T_i, K_j) - C_\text{mkt}^\text{ask}(T_i, K_j)}{C_\text{mkt}^\text{ask}(T_i, K_j)}, 0 \right\} + \text{max} \left\{\frac{C_\text{mkt}^\text{bid}(T_i, K_j) - C_\Theta(T_i, K_j)}{C_\text{mkt}^\text{bid}(T_i, K_j)}, 0 \right\} \right)^2 \Bigg \}, \nonumber
\end{gather}
where $F_{\Theta}(T_i)$ denotes the model price of the  VIX futures expiring at $T_i$.  Similarly, $C_{\Theta}(T_i, K_j)$  represents the model call price for maturity $T_i$ and strike $K_j$. The parameters $\gamma^{T}, \omega_1^{T}, \omega_2^{T}$ as well as the initial variance $\xi_0^{T}$ are kept constant on the interval $[T_i, T_{i+\Delta}]$. Using the relative error approach adjusts for the differences between the prices of the derivatives. The weight $1/m$ adjusts for the difference in the number of quotes on the index (one futures versus many options). Similar objective functions are used in \cite{romer2022empirical, alfeus2020consistent, kokholm2015joint, chung2011information}. We use MATLAB's \texttt{fmincon}, with \texttt{TolFun = 1e-6}, \texttt{TolCon = 1e-6}, \texttt{Algorithm = sqp} and \texttt{MaxFunctionEvaluations = 5000}, to minimise the objective function.  

Numerical experiments for the accuracy of the pricing formulas in \cite{bourgey2023weak} select the mean-reversion parameter $k=1$ for use in the one-factor model. In our paper, we make the same choice of $k$ and calibrate the remaining parameters, $\gamma^T, \omega_1^T, \omega_2^T$, as well as the initial forward variance, $\xi_0^T$, to VIX futures and options. For the two-factor model, as stated in Section~\ref{sec:2} of this paper and in \cite{bergomi2015stochastic}, the parameters $k_1, k_2, \theta, \rho$ are chosen to match the term structure of the classical two-factor Bergomi model's vol-of-vol with a power-law benchmark (see \cite{bergomi2008smile, bergomi2015stochastic} for details); these parameters are not calibrated to market prices of VIX futures and options. For the purpose of calibrating VIX futures and options, we fix a realistic choice of these parameters--namely $k_1 = 7.54, k_2 = 0.24, \theta = 0.23, \rho = 0.7$ (Set III in \cite{bergomi2015stochastic})--and calibrate $\gamma^T, \omega_1^T, \omega_2^T$, $\xi_0^T$ to VIX futures and options. In our calibration experiments, the fixed parameters yield objective function values that are effectively zero, so changing them is unnecessary.  We, respectively, use $1000$ and $1450$ quantisation points in the one- and two-factor models.

It is well known that VIX smiles are upward sloping, and having a model that reproduces them is one of the reasons why Bergomi introduced the mixed two-factor model in \cite{bergomi2008smile}. In this section, we calculate the VIX smiles produced in our calibration. Using the model-generated VIX futures prices, we calculate VIX implied volatilities via Black's formula.  In Black's formula, for a given call price $C_0(T_i, K_j)$ with maturity $T_i$ and strike $K_j$, we determine the unique value of $\sigma(T_i, K_j)$ such that
\[
C_0 = e^{-r_iT_i}\left(F^{T_i}_0\mathcal{N}(d_1) - K_j\mathcal{N}(d_2) \right),
\]
where
\[
d_1 = \frac{\ln \left(F_0^{T_i}/K_j\right) + \left(\sigma^2/2\right)T_i}{\sigma\sqrt{T_i}}, \quad d_2 = d_1 -\sigma\sqrt{T_i},
\]
with $F_0^{T_i}$ representing the model VIX futures price, $r_i$ the risk-free interest rate, and $\mathcal{N}(\cdot)$ the cumulative distribution function of the standard normal distribution. 

\begin{figure}[h!]
    \centering
    \begin{minipage}{0.32\textwidth}
        \centering
        \includegraphics[width=\textwidth]{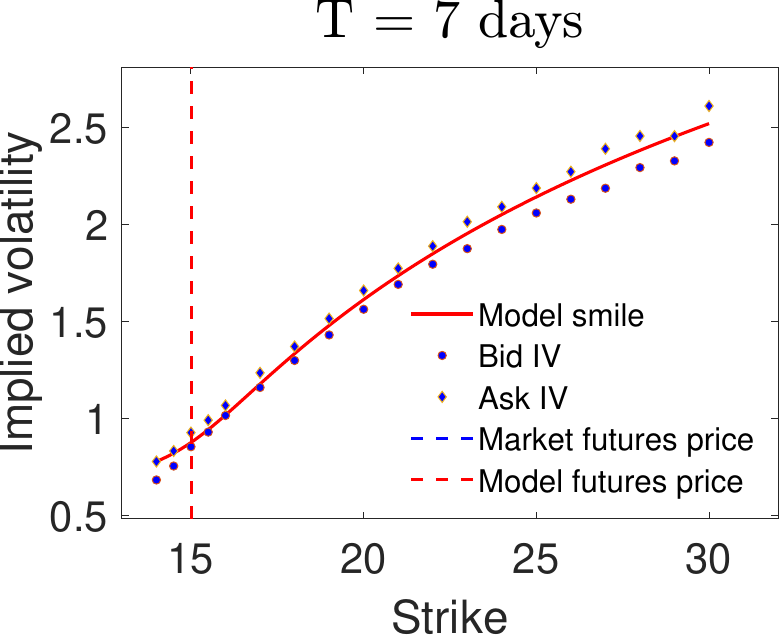}
    \end{minipage}
       \begin{minipage}{0.32\textwidth}
        \centering
        \includegraphics[width=\textwidth]{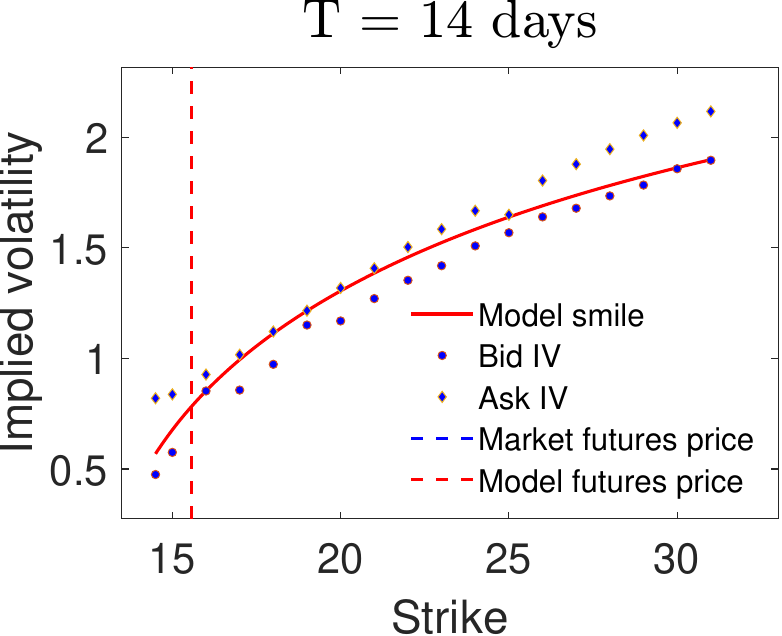}
    \end{minipage} 
    \begin{minipage}{0.32\textwidth}
        \centering
        \includegraphics[width=\textwidth]{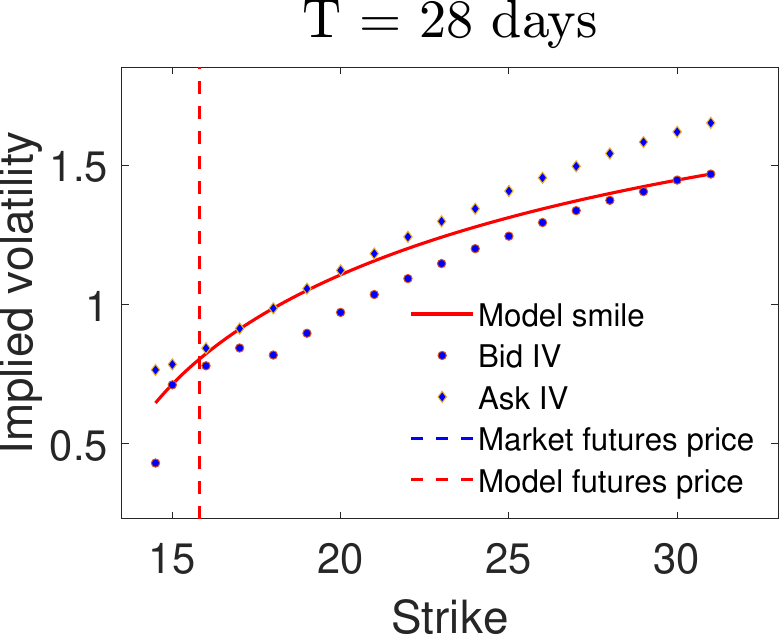}
    \end{minipage} \\
        \vspace{0.4cm} 
    \begin{minipage}{0.32\textwidth}
        \centering
        \includegraphics[width=\textwidth]{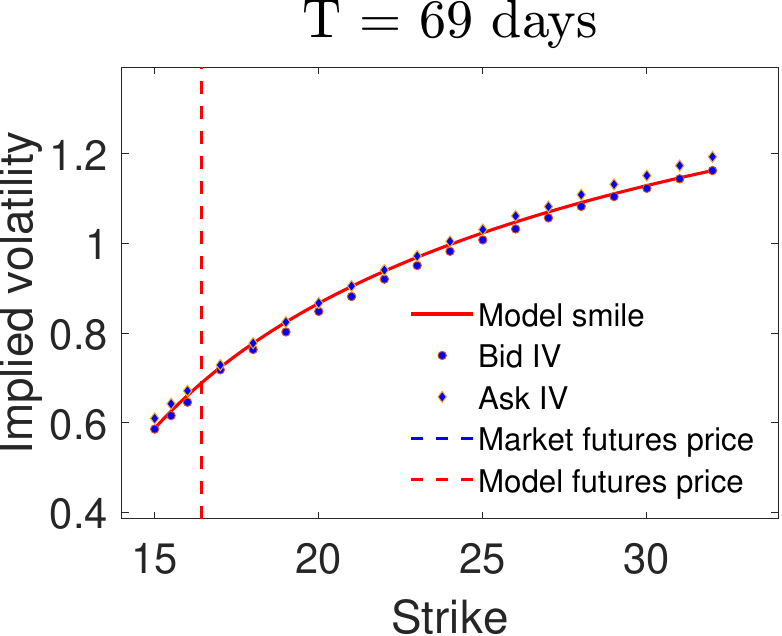}
    \end{minipage}
    \begin{minipage}{0.32\textwidth}
        \centering
        \includegraphics[width=\textwidth]{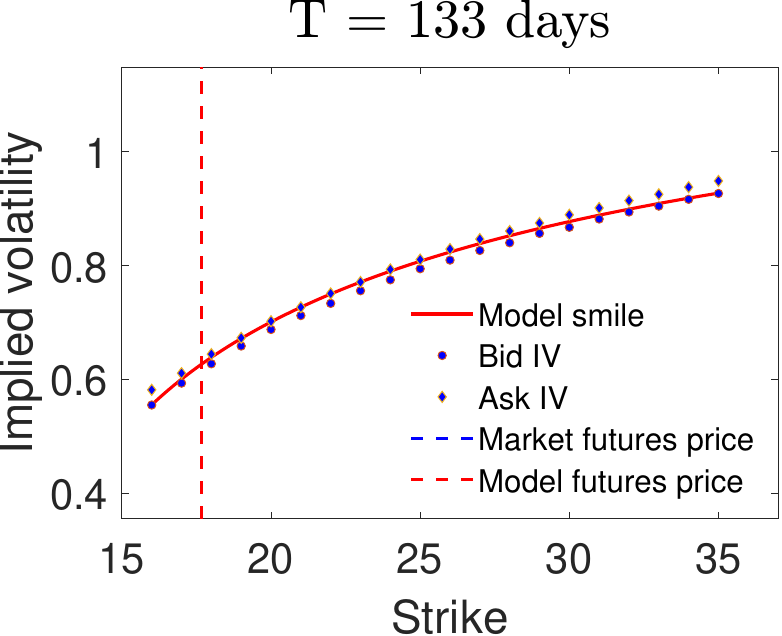}
    \end{minipage}
       \begin{minipage}{0.32\textwidth}
        \centering
        \includegraphics[width=\textwidth]{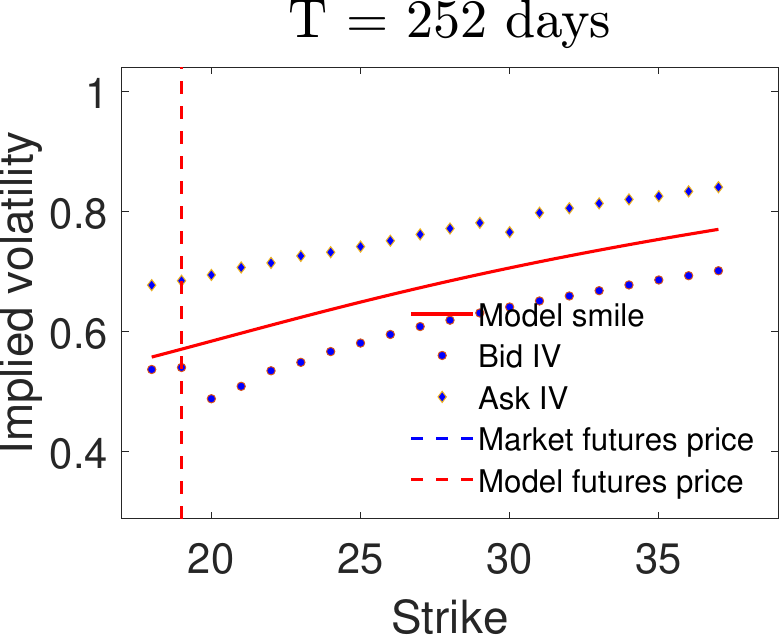}
    \end{minipage} 
    \caption{Futures (vertical lines) and smiles from the joint calibration to VIX futures and calls as of $09$ April  $2024$, using the one-factor model.}
     \label{fig:bid-ask-one-factor-9April} 
\end{figure}

\begin{figure}[h!]
    \centering
    \begin{minipage}{0.32\textwidth}
        \centering
        \includegraphics[width=\textwidth]{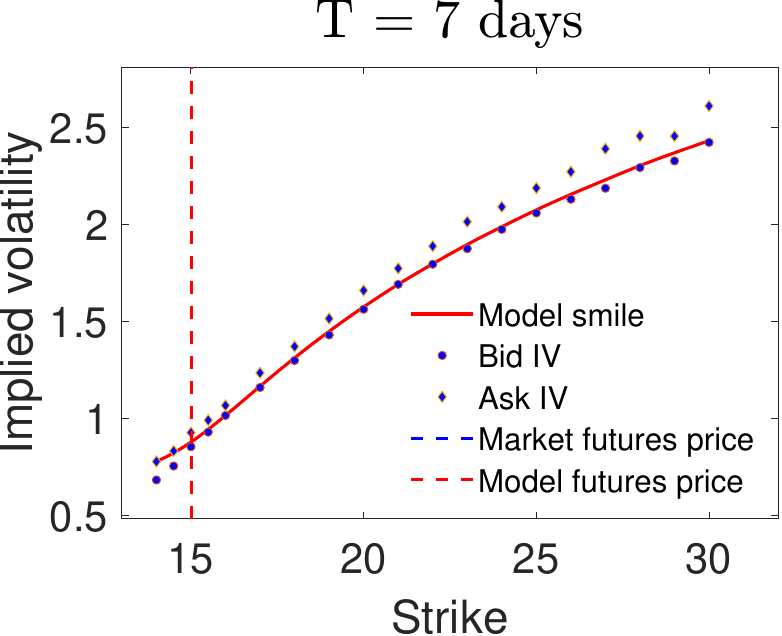}
    \end{minipage}
       \begin{minipage}{0.32\textwidth}
        \centering
        \includegraphics[width=\textwidth]{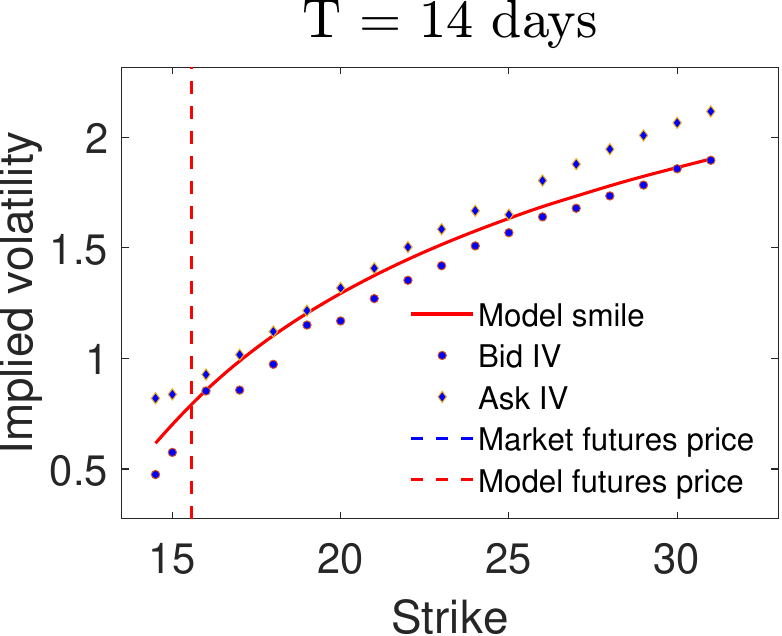}
    \end{minipage} 
    \begin{minipage}{0.32\textwidth}
        \centering
        \includegraphics[width=\textwidth]{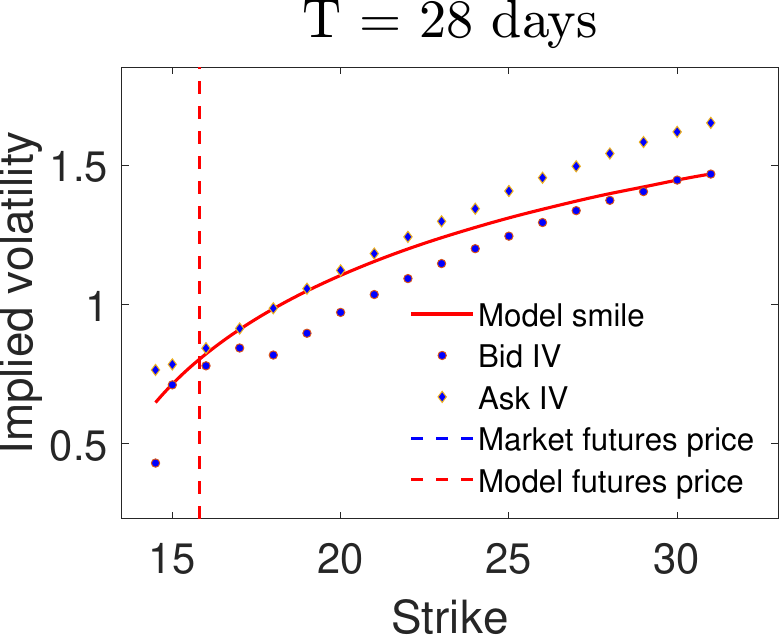}
    \end{minipage} \\
        \vspace{0.4cm} 
    \begin{minipage}{0.32\textwidth}
        \centering
        \includegraphics[width=\textwidth]{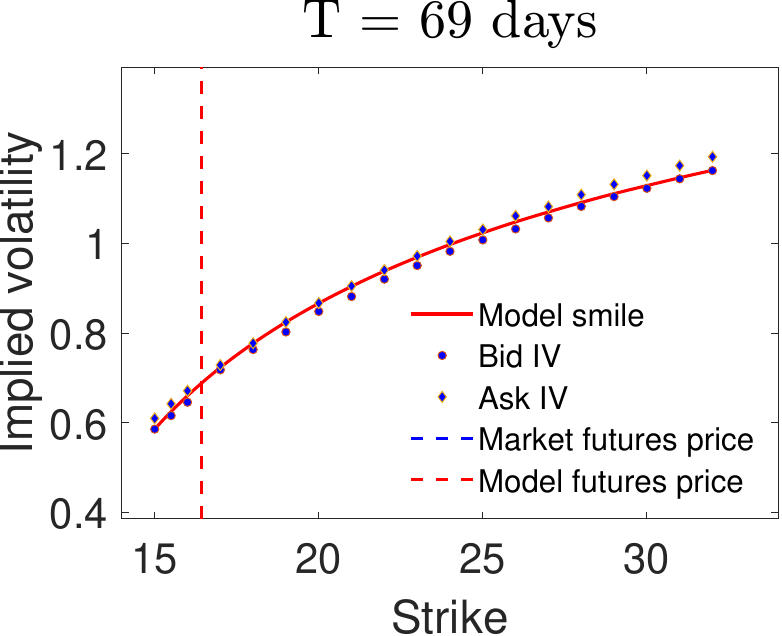}
    \end{minipage}
    \begin{minipage}{0.32\textwidth}
        \centering
        \includegraphics[width=\textwidth]{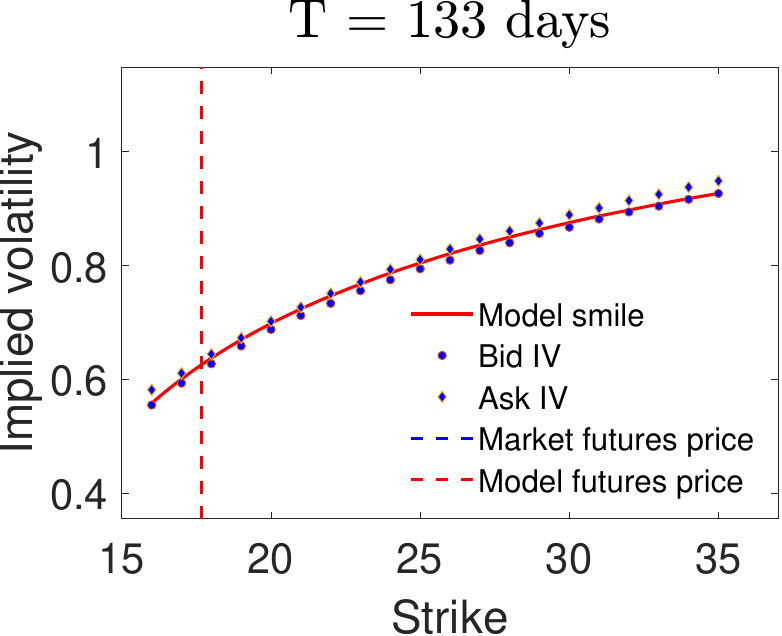}
    \end{minipage}
       \begin{minipage}{0.32\textwidth}
        \centering
        \includegraphics[width=\textwidth]{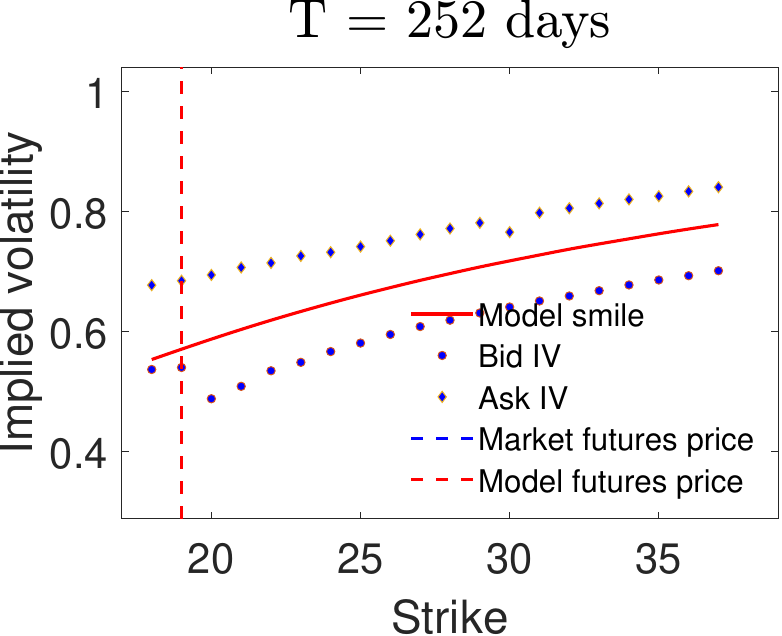}
    \end{minipage} 
    \caption{Futures (vertical lines) and smiles from the joint calibration to VIX futures and calls as of $09$ April  $2024$, using the two-factor model.}
     \label{fig:bid-ask-two-factor-9April} 
\end{figure}

\begin{figure}[h!]
    \centering
    \begin{minipage}{0.48\textwidth}
        \centering
        \includegraphics[width=\textwidth]{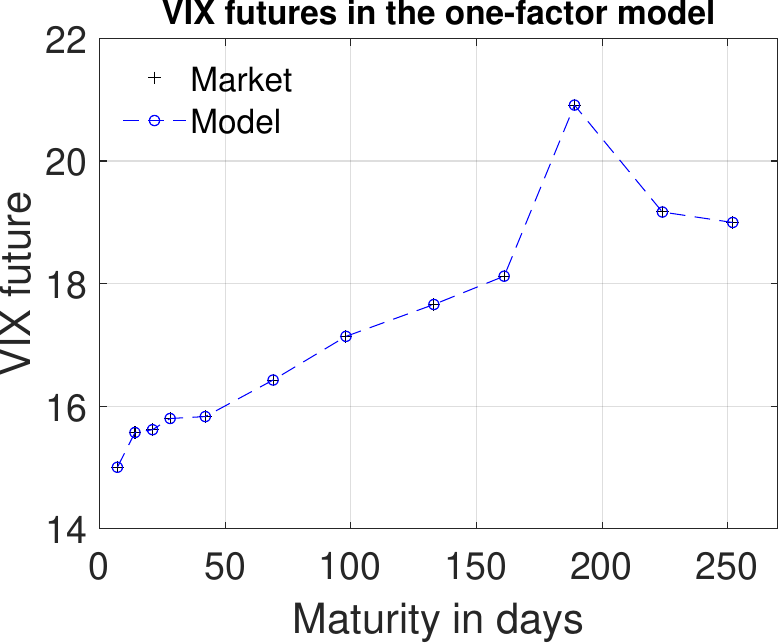}
    \end{minipage}
     \hspace{0.3cm}
    \begin{minipage}{0.48\textwidth}
        \centering
        \includegraphics[width=\textwidth]{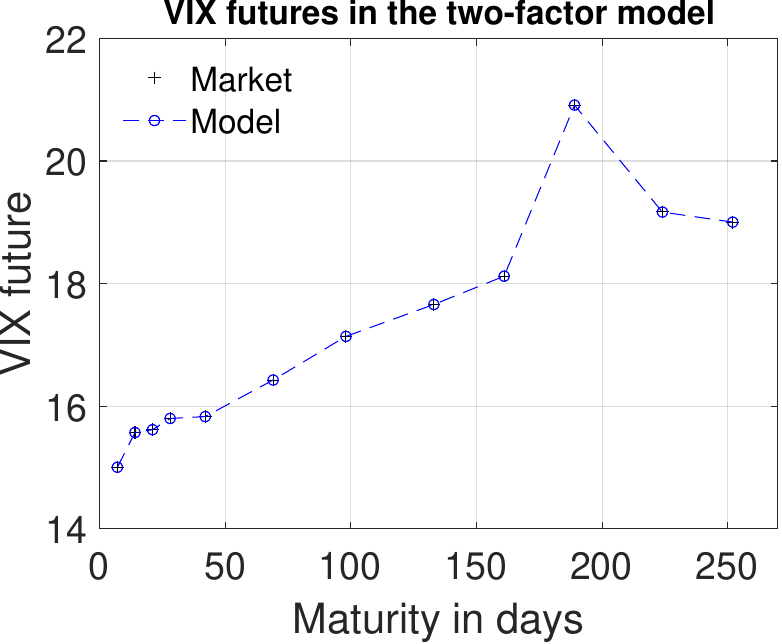}
    \end{minipage}
    \caption{VIX futures term structure from the joint calibration to VIX futures and calls as of $09$ April $2024$. \textit{Left:} Mixed one-factor Bergomi model. \textit{Right:} Mixed two-factor Bergomi model. }
     \label{fig:term-structure-april} 
\end{figure}

In Figures~\ref{fig:bid-ask-one-factor-9April} and~\ref{fig:bid-ask-two-factor-9April}, we present VIX smiles as obtained from our joint calibration to prices of \(09\) April, \(2024\), in the one- and two-factor models, respectively. As seen, models' smiles comfortably stay within the bid-ask implied volatilities. Model VIX futures prices match the market prices exactly.  Figure~\ref{fig:term-structure-april} shows the term structure of the VIX futures as of \(09\) April, \(2024\) in both models. It's evident from Figure~\ref{fig:term-structure-april} that the models' futures exactly match those of the market for the entire daily surface as of \(09\) April, \(2024\). Notably, the two models are indistinguishable by looking at the plots.

Table~\ref{tab:parameters-april} shows the models' parameters coming from the joint calibration as of $09$ April $2024$. Section~\ref{sec:parameter-stability} explores how the parameters change within and across daily surfaces.  

\begin{table}[h!]
    \centering
        \begin{threeparttable}
    \caption{VIX market-calibrated parameters for mixed Bergomi models as of  \(09\) April, \(2024\). }
    \label{tab:parameters-april}
    \begin{tabular}{c c c c c c}
        \toprule
        \makecell{Maturity \\ (days)} & Model & $\gamma^T$ & $\omega_1^T$ & $\omega_2^T$ & $\xi_0^T$ \\
        \midrule
        \multirow{2}{*}{$7$} & One-factor & $0.9154$ & $17.9773$&$1.2834$ & $2.3384\times 10^{-2}$ \\
         & Two-factor  & $0.9099$ & $23.9030$ & $1.6242$ & $2.3534\times 10^{-2}$ \\
        \midrule 
        \multirow{2}{*}{$14$} & One-factor  & $0.7992$ & $10.1150$ & $0.5242$ & $2.5717\times 10^{-2}$ \\ 
         & Two-factor  & $0.7529$ & $11.9445$ & $0.2027$& $ 2.5687\times 10^{-2}$ \\
        \midrule 
                \multirow{2}{*}{$21$} & One-factor & $0.7475$ & $8.8105$ & $0.3626$ & $ 2.6586\times 10^{-2}$ \\ 
         & Two-factor  & $0.7179$ & $11.5348$ & $0.0255$ & $2.6585\times 10^{-2}$ \\
        \midrule
                \multirow{2}{*}{$28$} & One-factor  & $ 0.6070$ & $5.8838$ & $0.0028$ & $2.7367\times 10^{-2}$ \\ 
         & Two-factor  & $0.6289$ & $8.7869$ & $0.2092$ & $2.7548\times 10^{-2}$ \\
        \midrule
                \multirow{2}{*}{$42$} & One-factor  & $0.7150$ & $8.1597$ & $ 0.8695$ & $ 3.0049 \times 10^{-2}$ \\ 
         & Two-factor  & $0.6548$ & $9.8553$ & $0.7096$ & $2.9108\times 10^{-2}$ \\
        \midrule 
                \multirow{2}{*}{$69$} & One-factor & $0.5495$ & $5.0842$ & $0.2774$ & $3.2823\times 10^{-2}$ \\ 
         & Two-factor  & $0.5765$ & $9.0625$& $0.7617$& $3.3348\times 10^{-2}$ \\
        \midrule 
                \multirow{2}{*}{$98$} & One-factor  & $0.4878$ & $4.4036$ & $0.2664$ & $3.7267\times 10^{-2}$ \\
         & Two-factor  & $0.4862$ & $7.8423$ & $0.5295$& $3.7567\times 10^{-2}$ \\
        \midrule 
                \multirow{2}{*}{$133$} & One-factor  & $0.4046$ & $3.6996$ & $0.0934$ & $4.0503\times 10^{-2}$\\ 
         & Two-factor  & $0.3920$ & $6.8552$ & $0.0991$ & $4.0660 \times 10^{-2}$ \\
        \midrule 
                \multirow{2}{*}{$161$} & One-factor  & $0.3601$ & $3.4183$ & $0.0008$ & $4.3517\times 10^{-2}$ \\ 
         & Two-factor  & $0.3465$ & $6.4943$ & $0$ & $4.3725\times 10^{-2}$ \\
        \midrule 
                \multirow{2}{*}{$189$} & One-factor & $0.5348$ & $5.0020$ & $0.9673$ & $6.7826 \times 10^{-2}$ \\ 
         & Two-factor  & $0.3176$ & $6.3029$ & $0.0016$ & $5.9497\times 10^{-2}$ \\
        \midrule 
                \multirow{2}{*}{$224$} & One-factor  & $0.5112$ & $5.5986$ & $1.2783$& $ 6.7825\times 10^{-2}$ \\
         & Two-factor  & $0.4593$ & $8.7608$ & $1.8762$ & $5.9483 \times 10^{-2}$\\
        \midrule 
                \multirow{2}{*}{$252$} & One-factor  & $0.5136$ & $5.5985$ & $1.2779$ & $6.7825\times 10^{-2}$ \\ 
         & Two-factor  & $0.4560$ & $8.7610$ & $1.8765$ & $5.9482 \times 10^{-2}$ \\
        \bottomrule 
    \end{tabular}
    \end{threeparttable}
\end{table}
Tables~\ref{tab:futures-statistics}, \ref{tab:futures-calls}, and~\ref{tab:futures-calls-pointwise} show the global surface calibration statistics for the models; we use these statistics to evaluate the individual and relative static model performance. We summarise the calibration statistics using the calibration performance metrics: relative error (RE) for the futures and average relative bid-ask error for the calls (ARBAE),  both calculated at the slice level;  and relative bid-ask error (RBAE) for the calls, calculated for each call price in the global surface.  

 We calculate RE for each fit to VIX futures using 
\begin{align}\label{eqn:are-futures}
    RE =  \frac{\left|F_\text{mkt}(T_i) - F_\Theta(T_i)\right|}{F_\text{mkt}(T_i)}.
\end{align}

ARBAE measures the average relative error of model prices outside the bid-ask bounds for each fit. For each maturity of VIX calls, we calculate ARBAE of the corresponding fit as follows: 
\begin{align}\label{eqn:arbae-calls}
  ARBAE = & \frac{1}{m}\sum_{j=1}^m \Bigg( \text{max} \left\{ \frac{C_\Theta(T_i, K_j) - C_\text{mkt}^{\text{ask}}(T_i, K_j)}{C_\text{mkt}^{\text{ask}}(T_i, K_j)}, 0\right\} + \nonumber \\ 
  & \text{max} \left\{ \frac{C^{\text{bid}}_\text{mkt}(T_i, K_j) - C_\Theta(T_i, K_j)}{C^{\text{bid}}_\text{mkt}(T_i, K_j)}, 0\right\}\Bigg). 
\end{align}
where $m$ is the total number of calls for that maturity. 

RBAE measures the relative error of model prices outside the bid-ask corridor, calculated for each call price. For each call price in the global fit, we calculate RBAE as follows: 
\begin{align}\label{eqn:rbae-calls}
    RBAE=   \text{max} \left\{ \frac{C^{i,j,l}_\Theta - C_\text{ask}^{i, j, l}}{C_\text{ask}^{i, j, l}}, 0\right\} +  \text{max} \left\{ \frac{C^{i,j,l}_\text{bid} - C_\Theta^{i, j, l}}{C_\text{bid}^{i, j, l}}, 0\right\}, 
\end{align}
where $C_\text{mkt}^{i,j, l}$ is the market VIX call price with maturity $T_i$, on trading day $l$, with strike $K_j$; $C_\Theta^{i,j,l}$ is the model equivalent of $C_\text{mkt}^{i,j, l}$.

In Tables~\ref{tab:futures-statistics}, \ref{tab:futures-calls} and~\ref{tab:futures-calls-pointwise}, \textbf{bold} indicates the better-performing model (i.e., the model with the lower error) for each statistical measure. While both models demonstrate remarkable performance, the two-factor model slightly outperforms the one-factor model overall. This slight advantage of the two-factor model is not unexpected. Generally, a two-factor model is expected to significantly outperform a one-factor model. However, in our parametrisation, both models feature three calibrated parameters, which explains why the performance improvement of the two-factor model over the one-factor model is only marginal. Nevertheless, the calibration power of each model is exceptional. 

\begin{table}[h!]
    \centering
    \begin{threeparttable}
        \caption{Summary statistics for the relative errors of VIX futures across all calibration dates. The table reports key metrics of relative errors, including the mean, standard deviation (SD), minimum, maximum, and the 95th and 99th percentiles. The lower error for each statistical measure is shown in \textbf{bold}. }
    \label{tab:futures-statistics}
    \begin{tabular}{l c c c c c c c }
    \toprule
       Model &  mean RE & SD  & min  & max  & $95\%$ & $99 \%$\\ \midrule
        One-factor & $3.54\times 10^{-5}$ & $0.000582$ & $0$ & $0.0170$  & $\bm{2.58\times 10^{-6}}$ & $0.000572$ \\ 
        Two-factor & $\bm{1.30\times 10^{-5}}$ & $\bm{0.000160}$  & $0$  & $\bm{0.00360}$  & $2.87\times 10^{-6}$ & $\bm{0.000250}$ \\
        \midrule
    \end{tabular}
        \end{threeparttable}
\end{table}

\begin{table}[h!]
    \centering
    \begin{threeparttable}
    \small 
        \caption{Summary statistics for the ARBAE of VIX calls across all calibration dates. The table reports key metrics of ARBAE, including the mean, standard deviation (SD), minimum, maximum, and the 95th and 99th percentiles. The lower error for each statistical measure is shown in \textbf{bold}. }
    \label{tab:futures-calls}
    \begin{tabular}{l c c c c c c   c }
    \toprule
        Model & mean ARBAE & SD  & min  & max  & $95\%$ & $99\%$ \\ \midrule 
          One-factor & $3.45\times 10^{-5}$  &  $0.000395$ & $0$ & $0.00762$   & $\bm{6.68\times 10^{-7}}$  & $0.000682$\\
        Two-factor & $\bm{2.84\times 10^{-5}}$ & $\bm{0.000351}$ & $0$ & $0.00762$  & $7.10\times 10^{-7}$   & $\bm{0.000525}$\\
        \bottomrule
    \end{tabular}
        \end{threeparttable}
\end{table}

\begin{table}[h!]
    \centering
    \begin{threeparttable}
       \caption{Summary statistics for the RBAE of VIX calls across all calibration dates. The table reports key metrics of RBAE, including the mean, standard deviation (SD), minimum, maximum, and the 95th and 99th percentiles. The lower error for each statistical measure is shown in \textbf{bold}. }
    \label{tab:futures-calls-pointwise}
    \begin{tabular}{l c c c c c c   c }
    \toprule
        Model & mean RBAE & SD  & min  & max  & $95\%$ & $99\%$ \\ \midrule 
          One-factor & $3.09\times 10^{-5}$  &  $0.000813$ & $0$ & $0.0523$   & $0$  & $2.53 \times 10^{-6}$\\
        Two-factor & $\bm{2.46\times 10^{-5}}$ & $\bm{0.0007}$ & $0$ & $\bm{0.0507}$  & $0$   & $\bm{2.52\times 10^{-6}}$      \\
        \bottomrule
    \end{tabular}
        \end{threeparttable}
\end{table}

\begin{figure}[h!]
    \centering
    \begin{minipage}{0.48\textwidth}
        \centering
        \includegraphics[width=\textwidth]{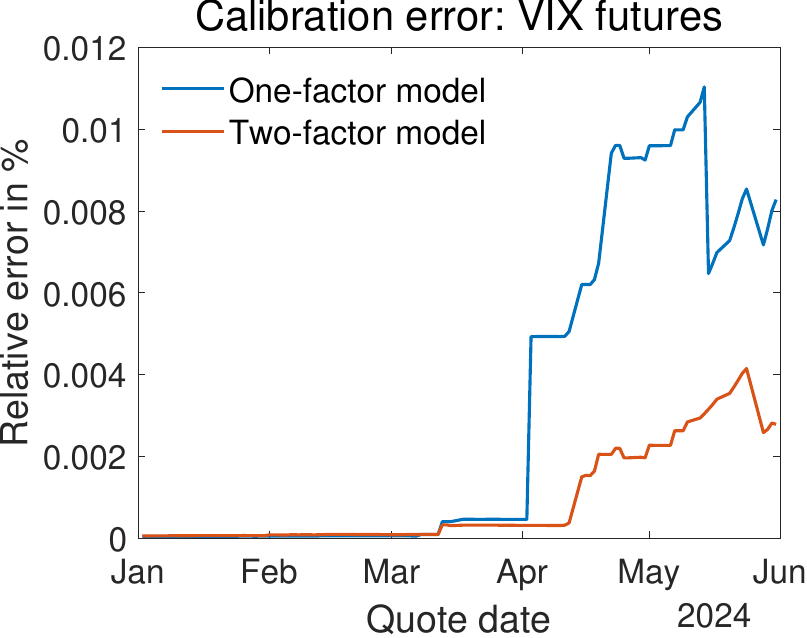}
    \end{minipage}
     \hspace{0.3cm}
    \begin{minipage}{0.48\textwidth}
        \centering
        \includegraphics[width=\textwidth]{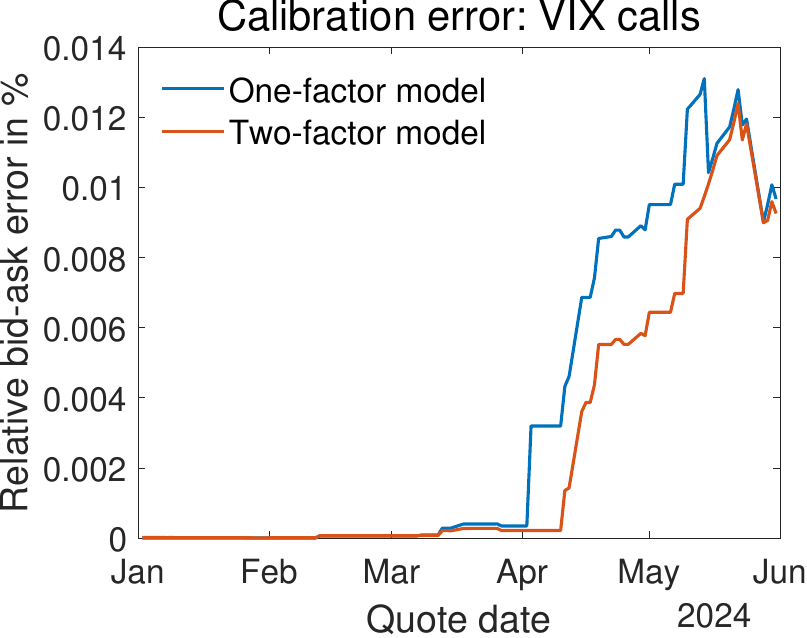}
    \end{minipage}
    \caption{Evolution of the $30$-day moving average calibration error for the joint calibration to VIX futures and calls. \textit{Left:} VIX futures. \textit{Right:} VIX calls. }
     \label{fig:global-calibration-error} 
\end{figure}

Figure~\ref{fig:global-calibration-error} presents the historical time series of daily average calibration errors for fits to futures and calls in the two models. Daily average errors are calculated via \eqref{eqn:are-futures} and~\eqref{eqn:arbae-calls}, and smoothed using $30$-day moving averages. The results indicate that the historical calibration errors are consistently low, with the two-factor model slightly outperforming the one-factor model overall. Up to mid-March 2024, the daily average calibration errors are almost zero. However, from mid-March to mid-May, calibration errors increase due to the high prices of VIX futures expiring in October 2024, driven by the U.S. elections conducted on 5 November 2024. As a result, models' fits deteriorate slightly to accommodate the rising VIX futures prices, with the one-factor model experiencing greater deterioration.  Nevertheless, the $30$-day moving averages of the errors remain below $1.5$ basis points.

\section{Stability of calibrated parameters}\label{sec:parameter-stability}
This section evaluates the stability of calibrated parameters by visualising their evolution over time and conducting numerical tests on their robustness.

The initial forward variance, $\xi_0^T$, will be treated in two ways: (1) as a market parameter and will be stripped from the market in this case, or (2) as a calibrated parameter, like in Section~\ref{sec:calibration}.  In practice, as a market parameter, $\xi_0^T$ can be obtained in two ways: (1) through SPX log-contract replication (see \cite{bergomi2015stochastic, jaber2022quintic, guyon2022does}), or (2) via replication using VIX futures, out-of-the-money (OTM) VIX calls, and OTM VIX puts (see \cite{bergomi2008smile, bergomi2015stochastic}). For the purposes of pricing VIX futures and calls, we replicate $\xi_0^T$ using the VIX market.  

\begin{figure}[!h]
\centering
\includegraphics[scale=0.8]{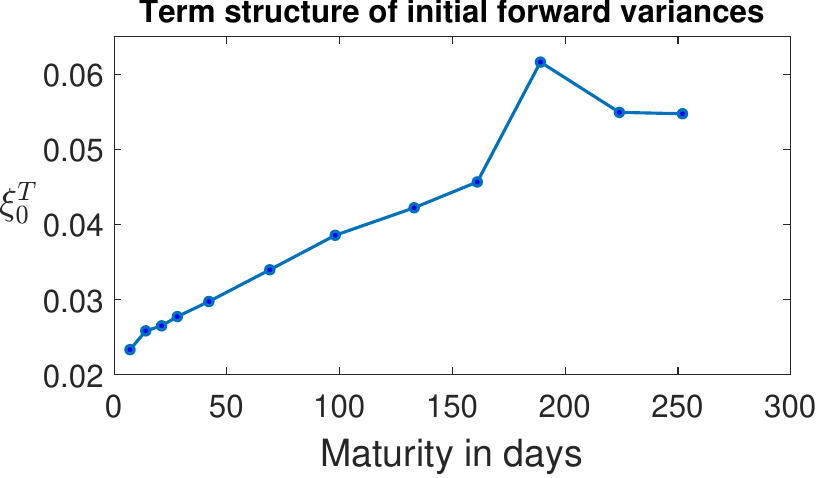}
	\caption{Term structure of initial forward variances as of  \(09\) April \(2024\).  The blue dots denote the market forward variance as obtained by the formula \eqref{eqn:initial-variance-term-structure}, in between linear interpolations are shown.} 
 \label{fig: initial variance term structure}
\end{figure}

For  constant $\xi_0^T$ over the interval $[T_i, T_i+\Delta]$, with $T\in [T_i, T_i+\Delta]$, the model-independent  equation relating $\xi_0^T$ to the VIX market is given by  
\begin{align}\label{eqn:initial-variance-term-structure}
    \xi_0^T = \left( F_0^{T_i} \right)^2 + 2\int_0^{F_0^{T_i}}\mathcal{P}\left(K, F_0^{T_i}\right)dK + 2\int^\infty_{F_0^{T_i}}\mathcal{C}\left(K, F_0^{T_i}\right)dK,
\end{align}  
where $F_0^{T_i}$ is the VIX futures for expiry $T_i$, observed at $t=0$, and $\mathcal{P}\left(K, F_0^{T_i}\right)$ and $\mathcal{C}\left(K, F_0^{T_i}\right)$ are the undiscounted market prices of put and call options on $F_0^{T_i}$, respectively. We apply~\eqref{eqn:initial-variance-term-structure} to market data, using all available put and call options alongside the corresponding VIX futures, on a slice-by-slice basis. The two integrals are computed using the trapezoidal rule. On each day, we obtain a term structure of initial variances. Figure~\ref{fig: initial variance term structure} illustrates this term structure as of \(09\) April \(2024\).

\subsection{Evolution of calibrated parameters}
To preserve the structure of the model as much as possible, the parameters whose time evolution we consider are $\gamma^T, \omega_1^T$ and $\omega_2^T$. The initial forward variance, $\xi_0^T$, will be recalculated daily using~\eqref{eqn:initial-variance-term-structure}. Consequently,  daily calibration focuses on the parameters $\gamma^T, \omega_1^T, \omega_2^T$.  Although using $\xi_0^T$ as stripped from the market by \eqref{eqn:initial-variance-term-structure} (as opposed to including it in the calibration) leads to some loss in accuracy, the model prices still match the market prices very well. The calibration accuracy of the models under this approach is shown in Tables \ref{tab:futures-statistics-market-initial-variance},  \ref{tab:futures-calls-market-initial-variance} and \ref{tab:futures-calls-pointwise-market-initial-variance}. As can be seen, the results in these tables exhibit reduced accuracy compared to those in Tables \ref{tab:futures-statistics}, \ref{tab:futures-calls} and \ref{tab:futures-calls-pointwise}, where $\xi_0^T$ is also calibrated to market data. This reduction in calibration power is mentioned in Section 7.7 of \cite{bergomi2015stochastic}, but we provide numerical results in contrast to their case. 

\begin{table}[h!]
    \centering
    \begin{threeparttable}
        \caption{Summary statistics for the relative errors of VIX futures across all calibration dates. The table reports key metrics of relative errors, including the mean, standard deviation (SD), minimum, maximum, and the 95th and 99th percentiles. }
    \label{tab:futures-statistics-market-initial-variance}
    \begin{tabular}{l c c c c c c c }
    \toprule
       Model  & mean RE & SD  &  min  & max  & $95\%$ & $99 \%$\\ \midrule 
        One-factor & $1.81\times 10^{-4}$ & $0.000898$ & $0$ & $0.0118$  & $9.52\times 10^{-4}$ & $0.00485$ \\ 
      Two-factor & $2.23\times 10^{-4}$ & $0.0011$  & $0$  & $0.00123$  & $1.22\times 10^{-3}$ & $0.00626$\\
        \bottomrule
    \end{tabular}
        \end{threeparttable}
\end{table}

\begin{table}[h!]
    \centering
    \begin{threeparttable}
    \small 
        \caption{Summary statistics for the ARBAE of VIX calls across all calibration dates. The table reports key metrics of ARBAE, including the mean, standard deviation (SD), minimum, maximum, and the 95th and 99th percentiles. }
    \label{tab:futures-calls-market-initial-variance}
    \begin{tabular}{l c c c c c c   c }
    \toprule 
        Model & mean ARBAE & SD  & min  & max  & $95\%$ & $99\%$ \\ \midrule
          One-factor & $9.08\times 10^{-5}$  &  $0.00051$ & $0$ & $0.00987$   & $5.0\times 10^{-4}$  & $0.00225$\\
        Two-factor & $7.56\times 10^{-5}$ & $ 0.00051$ & $0$ & $0.00992$  & $2.33\times 10^{-4}$   & $0.00174$\\
        \bottomrule
      
    \end{tabular}
        \end{threeparttable}
\end{table}

\begin{table}[h!]
    \centering
    \begin{threeparttable}
       \caption{Summary statistics for the RBAE of VIX calls across all calibration dates. The table reports key metrics of RBAE, including the mean, standard deviation (SD), minimum, maximum, and the 95th and 99th percentiles. }
    \label{tab:futures-calls-pointwise-market-initial-variance}
    \begin{tabular}{l c c c c c c   c }
    \toprule
        Model & mean RBAE & SD  & min  & max  & $95\%$ & $99\%$ \\ \midrule 
          One-factor & $8.33\times 10^{-5}$  &  $0.0011$ & $0$ & $0.0703$   & $0$  & $0.00221$\\
        Two-factor & $6.78\times 10^{-5}$ & $0.00111$ & $0$ & $0.0743$  & $0$   & $0.00081$\\
        \bottomrule
      
    \end{tabular}
        \end{threeparttable}
\end{table}

\begin{figure}[h!]
    \centering
    \begin{minipage}{0.32\textwidth}
        \centering
        \includegraphics[width=\textwidth]{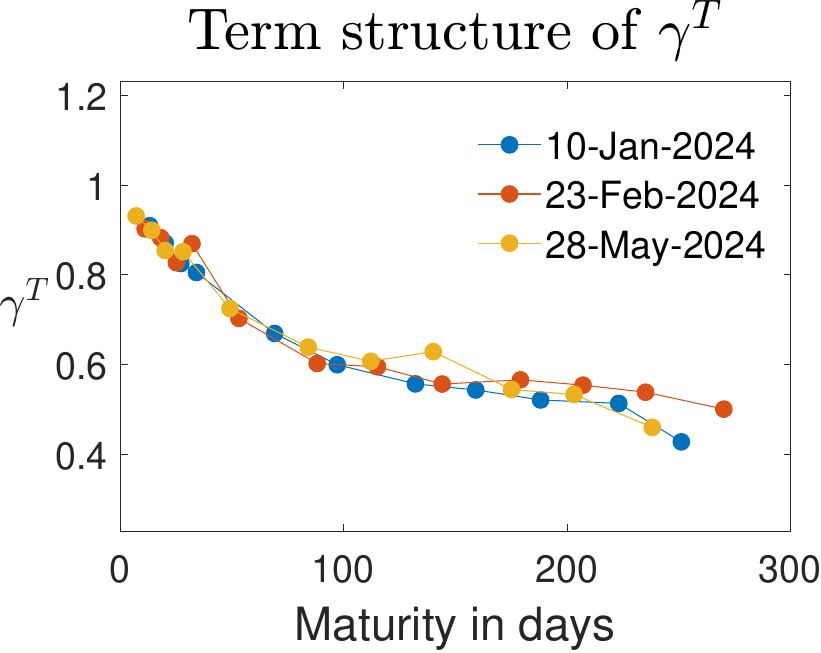}
    \end{minipage}
       \begin{minipage}{0.32\textwidth}
        \centering
        \includegraphics[width=\textwidth]{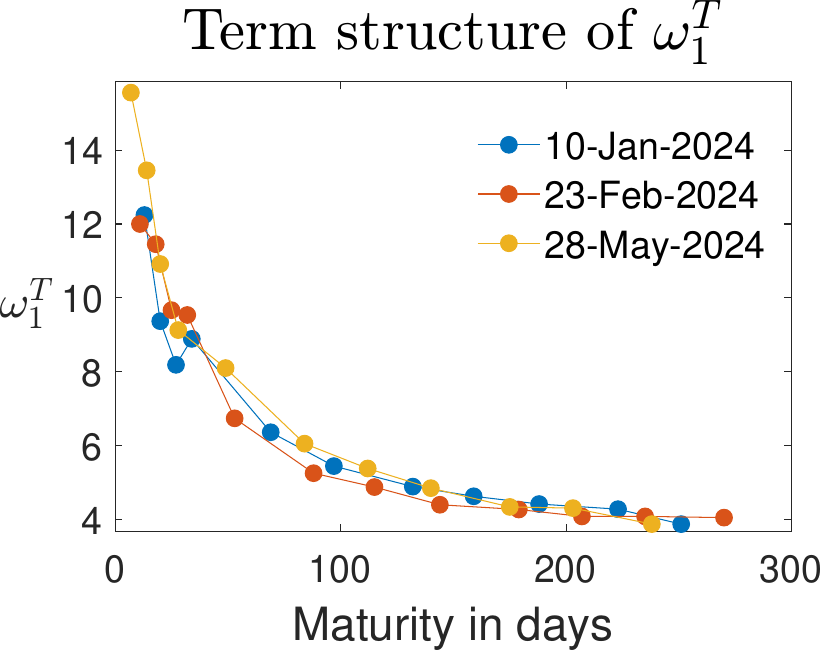}
    \end{minipage} 
    \begin{minipage}{0.32\textwidth}
        \centering
        \includegraphics[width=\textwidth]{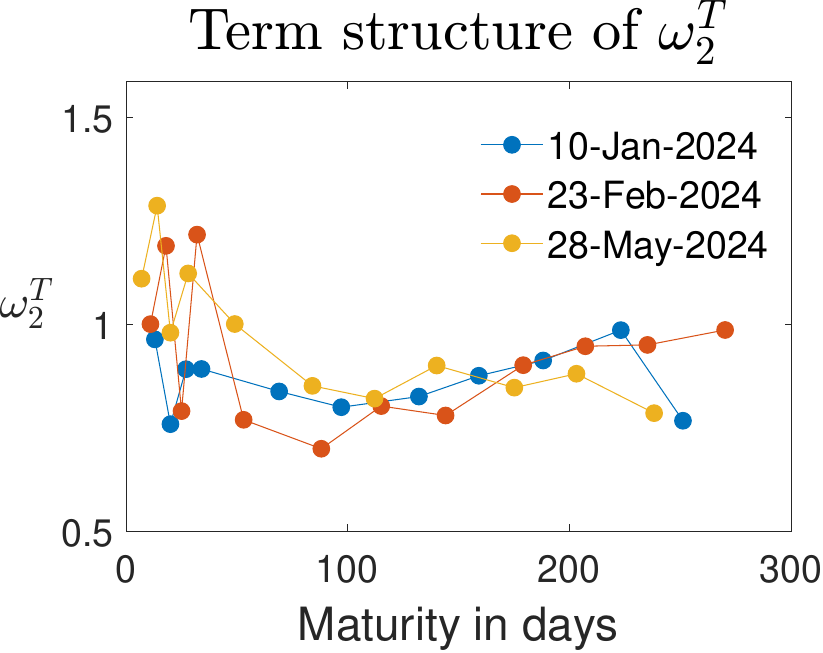}
    \end{minipage} 
    \caption{Term structures of calibrated parameters in the one-factor model.}
     \label{fig:parameter-term-structure-one-factor} 
\end{figure}

\begin{figure}[h!]
    \centering
    \begin{minipage}{0.32\textwidth}
        \centering
        \includegraphics[width=\textwidth]{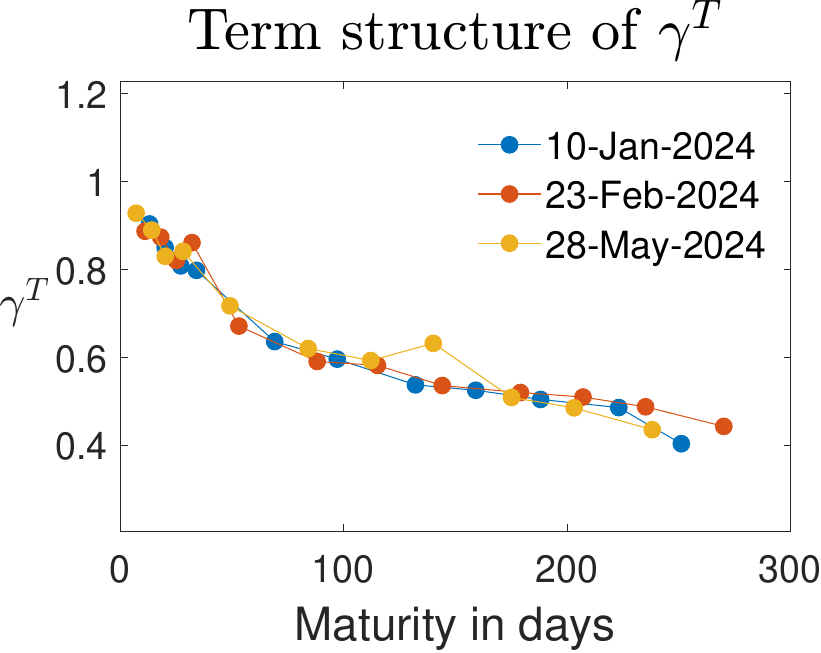}
    \end{minipage}
       \begin{minipage}{0.32\textwidth}
        \centering
        \includegraphics[width=\textwidth]{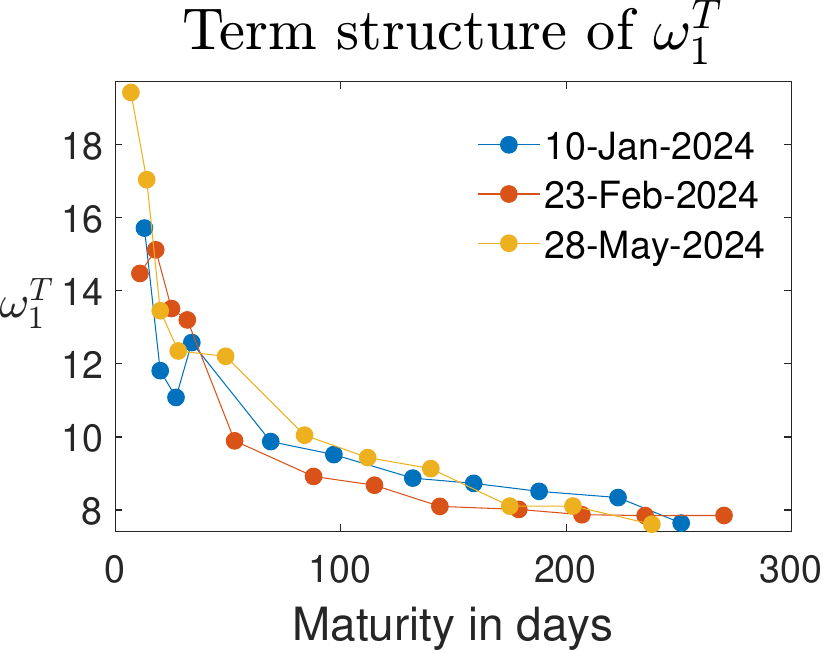}
    \end{minipage} 
    \begin{minipage}{0.32\textwidth}
        \centering
        \includegraphics[width=\textwidth]{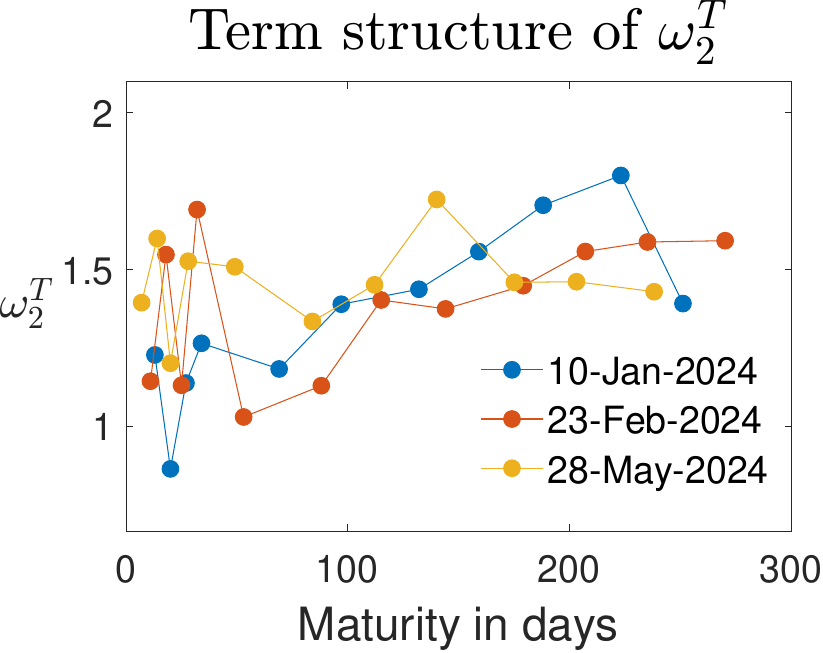}
    \end{minipage} 
    \caption{Term structures of calibrated parameters in the two-factor model.}
     \label{fig:parameter-term-structure-two-factor} 
\end{figure}

\begin{figure}[h!]
    \centering
    \begin{minipage}{0.32\textwidth}
        \centering
        \includegraphics[width=\textwidth]{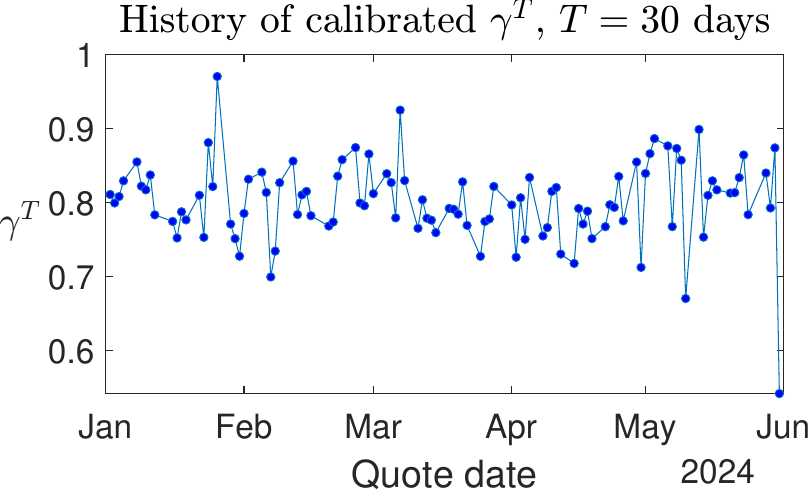}
    \end{minipage}
       \begin{minipage}{0.32\textwidth}
        \centering
         \includegraphics[width=\textwidth]{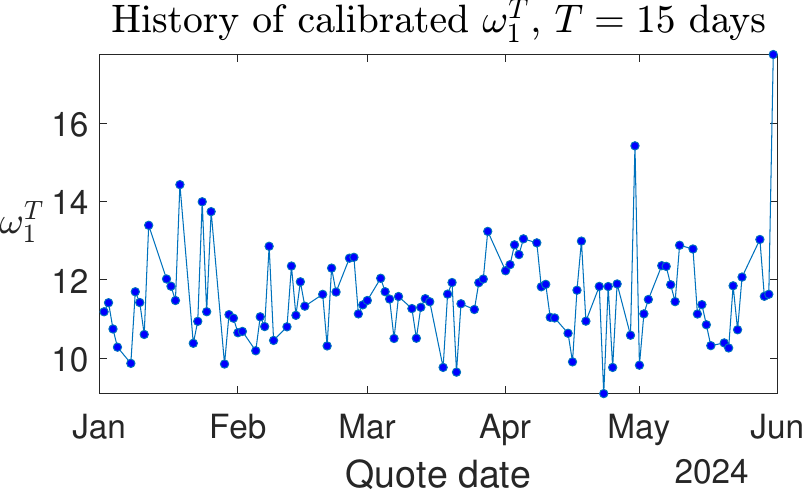}
    \end{minipage} 
    \begin{minipage}{0.32\textwidth}
        \centering
         \includegraphics[width=\textwidth]{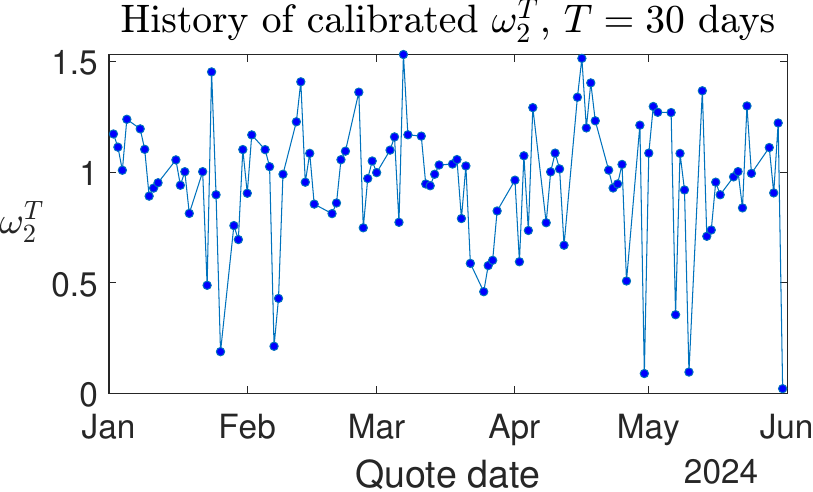}
    \end{minipage} \\ \vspace{0.3cm}
        \begin{minipage}{0.32\textwidth}
        \centering
         \includegraphics[width=\textwidth]{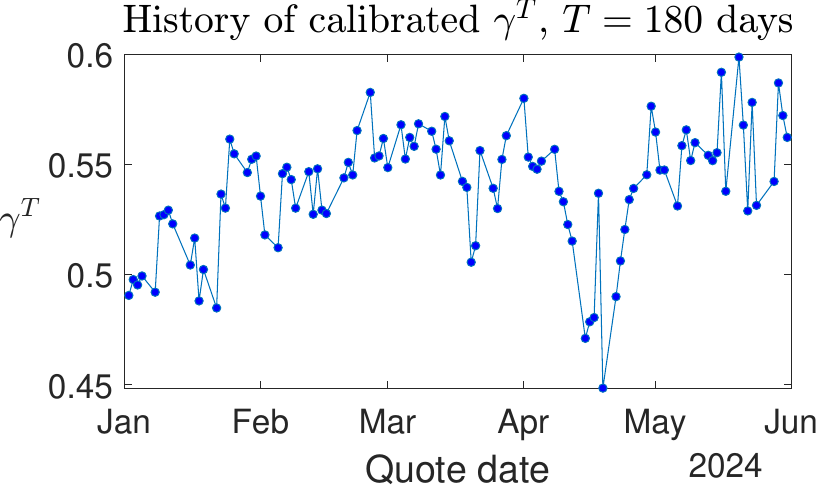}
    \end{minipage}
       \begin{minipage}{0.32\textwidth}
        \centering
        \includegraphics[width=\textwidth]{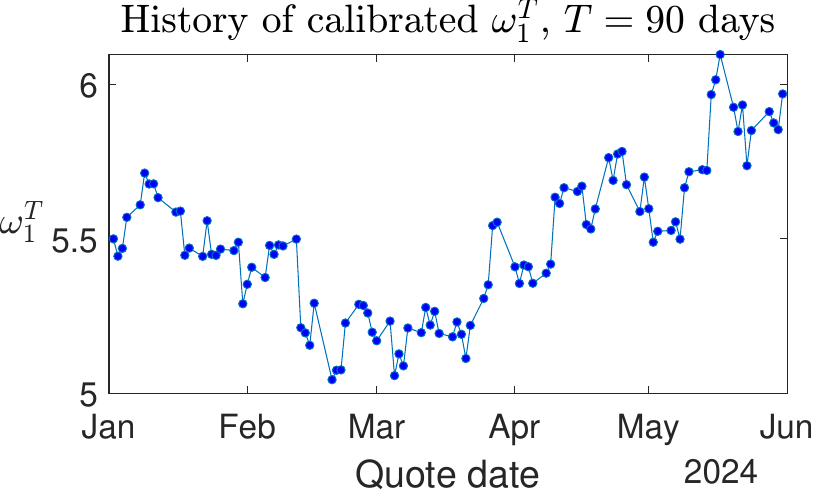}
    \end{minipage} 
    \begin{minipage}{0.32\textwidth}
        \centering
         \includegraphics[width=\textwidth]{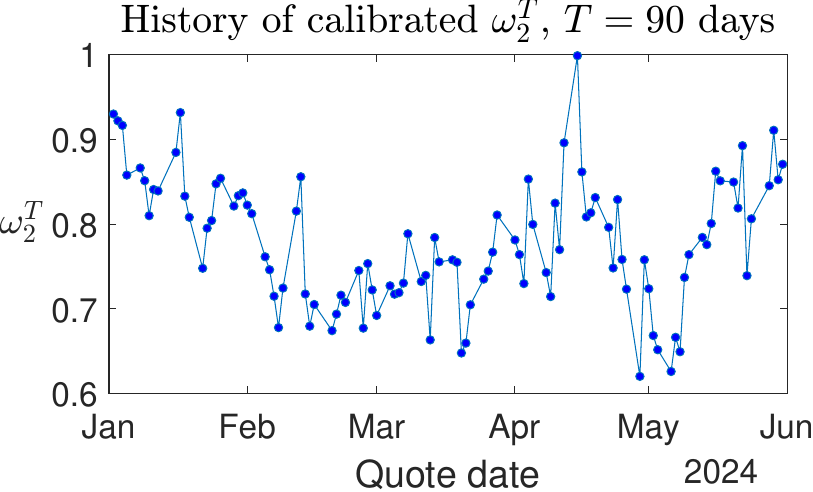}
    \end{minipage} 
    \caption{Evolution of calibrated parameters at different maturities in the one-factor model.}
     \label{fig:parameters-stability-across-days-one-factor} 
\end{figure}

\begin{figure}[h!]
    \centering
    \begin{minipage}{0.32\textwidth}
        \centering
        \includegraphics[width=\textwidth]{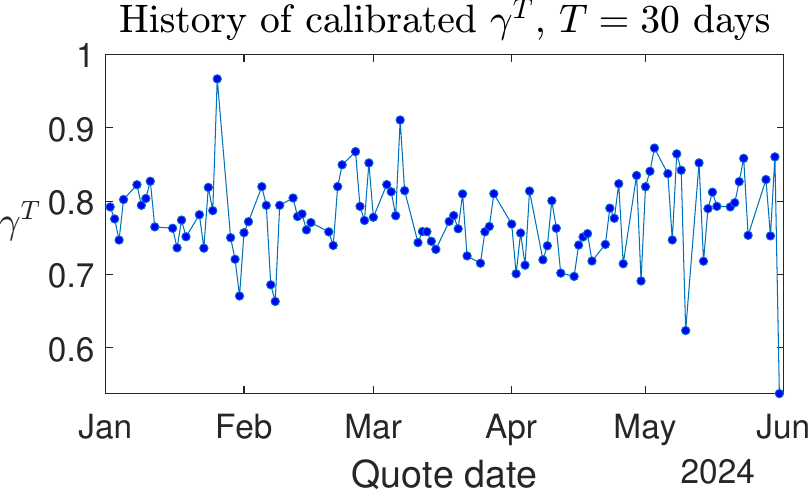}
    \end{minipage}
       \begin{minipage}{0.32\textwidth}
        \centering
         \includegraphics[width=\textwidth]{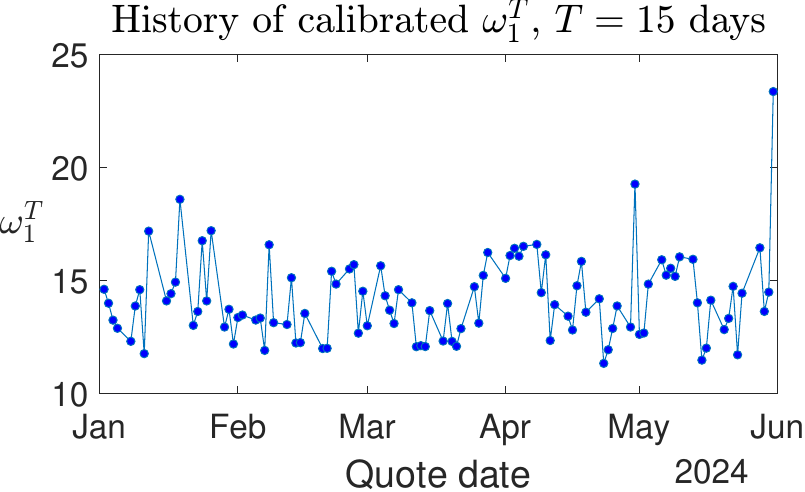}
    \end{minipage} 
    \begin{minipage}{0.32\textwidth}
        \centering
         \includegraphics[width=\textwidth]{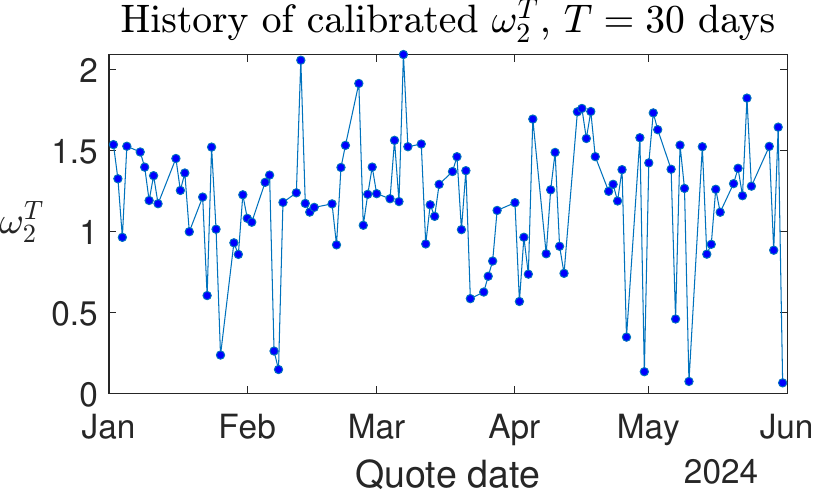}
    \end{minipage} \\ \vspace{0.3cm}
        \begin{minipage}{0.32\textwidth}
        \centering
         \includegraphics[width=\textwidth]{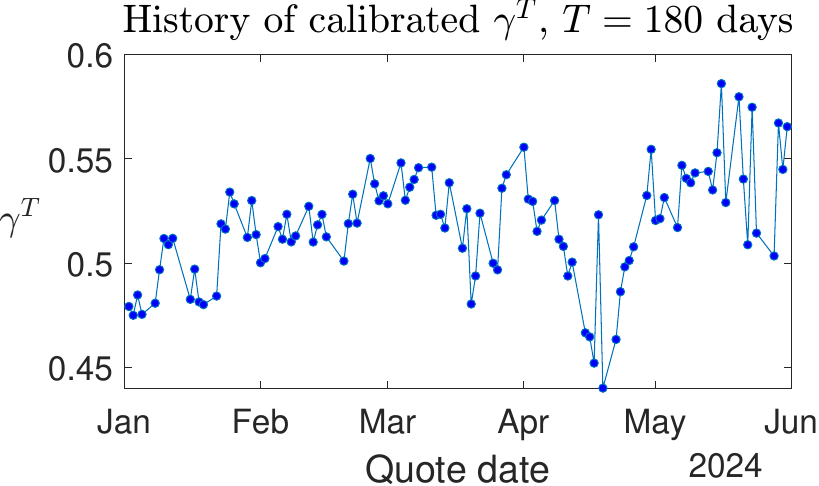}
    \end{minipage}
       \begin{minipage}{0.32\textwidth}
        \centering
        \includegraphics[width=\textwidth]{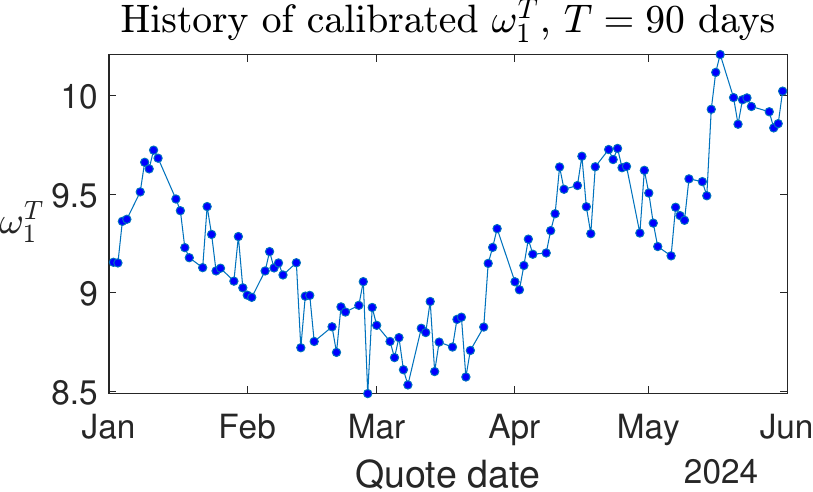}
    \end{minipage} 
    \begin{minipage}{0.32\textwidth}
        \centering
         \includegraphics[width=\textwidth]{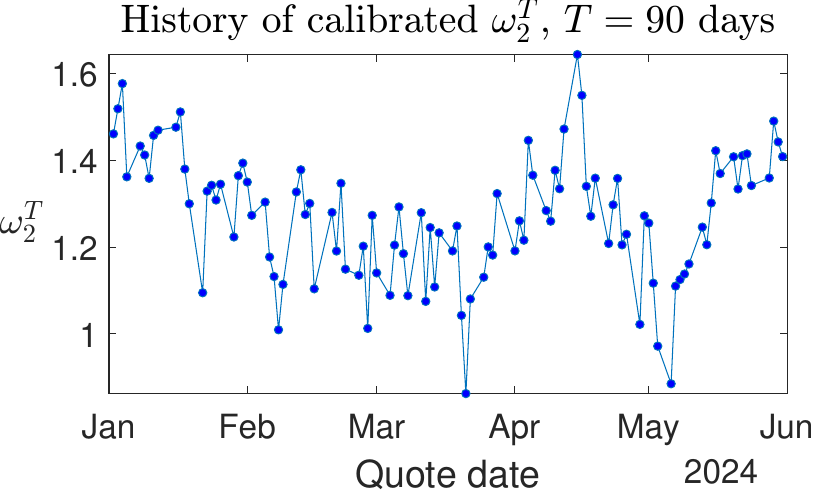}
    \end{minipage} 
    \caption{Evolution of calibrated parameters at different maturities in the two-factor model.}
     \label{fig:parameters-stability-across-days-two-factor} 
\end{figure}

Figures~\ref{fig:parameter-term-structure-one-factor} and~\ref{fig:parameter-term-structure-two-factor} illustrate the variation of calibrated parameters within a daily surface for a few selected days. All calibrated parameters show significant changes within the daily surfaces. 
Figures~\ref{fig:parameters-stability-across-days-one-factor} and~\ref{fig:parameters-stability-across-days-two-factor} show the historical progression of calibrated parameters, linearly interpolated across specified maturities throughout the study period. The parameters exhibit notable day-to-day variations, with similar behaviour observed across both models.

\subsection{Numerical test on parameter stability}
Here, we conduct numerical tests on parameter stability by assessing the pricing accuracy of the models under fixed parameters. A robust parametric model does not require frequent recalibration; a change in state variables alone should track changes in market prices. Thus, the pricing accuracy of a model under fixed parameters can be seen as a test on the stability of calibrated parameters.

To assess the pricing performance of the models while fixing different parameter sets, we perform the following tests: 
\begin{itemize}
    \item \textbf{Test 1:} In this test, we calibrate the parameters $\gamma^T$, $\omega_1^T$, and $\omega_2^T$ daily. After calibration, we keep these values fixed for the next $30$ days while computing VIX futures and call prices daily using the models. The initial forward variance, $\xi_0^T$, is recalculated daily via~\eqref{eqn:initial-variance-term-structure} both during calibration and during pricing under fixed parameters.  
    \item \textbf{Test 2:} Here, we calibrate $\gamma^T$, $\omega_1^T$, $\omega_2^T$, and $\xi_0^T$ daily, as described in Section~\ref{sec:calibration}, to achieve high calibration accuracy. Then, to compute daily model prices, we fix  $\gamma^T$, $\omega_1^T$, and $\omega_2^T$ for the next \(30\) trading days while refreshing the daily $\xi_0^T$ as stripped from the market via~\eqref{eqn:initial-variance-term-structure}, i.e., the difference to Test 1 lies in the initial calibration of $\xi_0^T$, while the updating of $\xi_0^T$ proceeds in the same way in Tests 1 and 2. 
    \item \textbf{Test 3:} In this test, we calibrate $\gamma^T$, $\omega_1^T$, $\omega_2^T$, and $\xi_0^T$ daily like in Section~\ref{sec:calibration}. After calibration, we fix  $\gamma^T$, $\omega_1^T$, and $\omega_2^T$ for the next \(30\) trading days while recalibrating $\xi_0^T$ to market prices daily, and compute the daily VIX future and call prices. 
    \item \textbf{Test 4:} Lastly,  we calibrate $\gamma^T$, $\omega_1^T$, $\omega_2^T$, and $\xi_0^T$ daily to market prices, keep their values fixed for the next \(30\) days while computing VIX futures and call prices. Thus, in this final test everything except the model state variables is held fixed.
\end{itemize}

\begin{figure}[h!]
    \centering
    \begin{minipage}{0.48\textwidth}
        \centering
        \includegraphics[width=\textwidth]{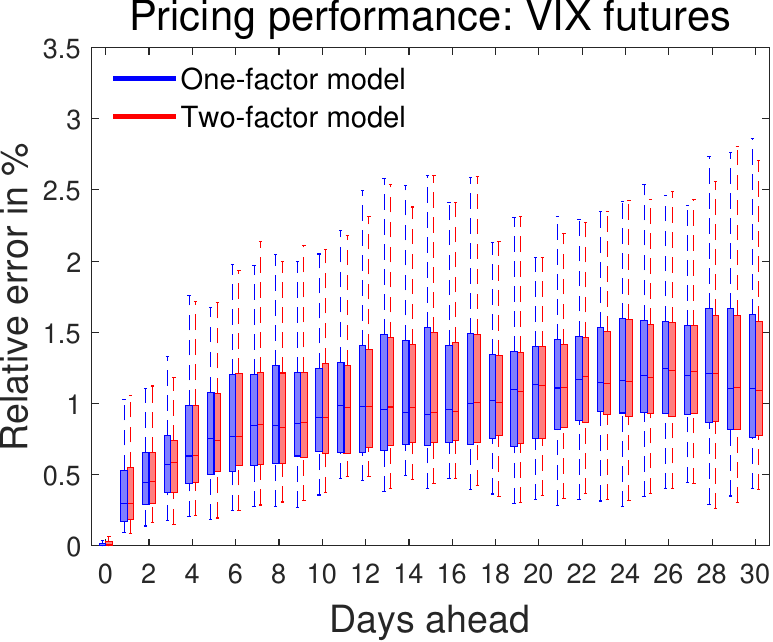}
    \end{minipage}
     \hspace{0.3cm}
    \begin{minipage}{0.48\textwidth}
        \centering
        \includegraphics[width=\textwidth]{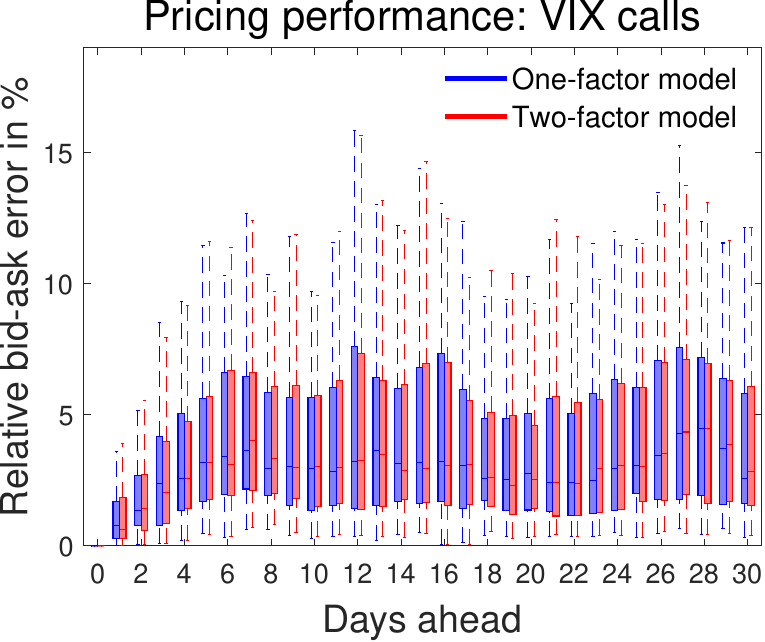}
    \end{minipage}\\
        \vspace{0.4cm} 
        \begin{minipage}{0.48\textwidth}
        \centering
        \includegraphics[width=\textwidth]{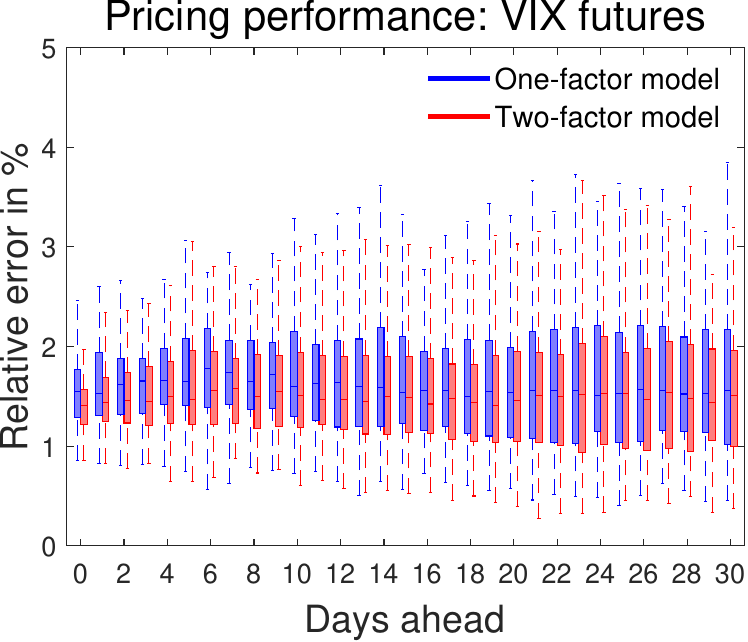}
    \end{minipage}
     \hspace{0.3cm}
    \begin{minipage}{0.48\textwidth}
        \centering
        \includegraphics[width=\textwidth]{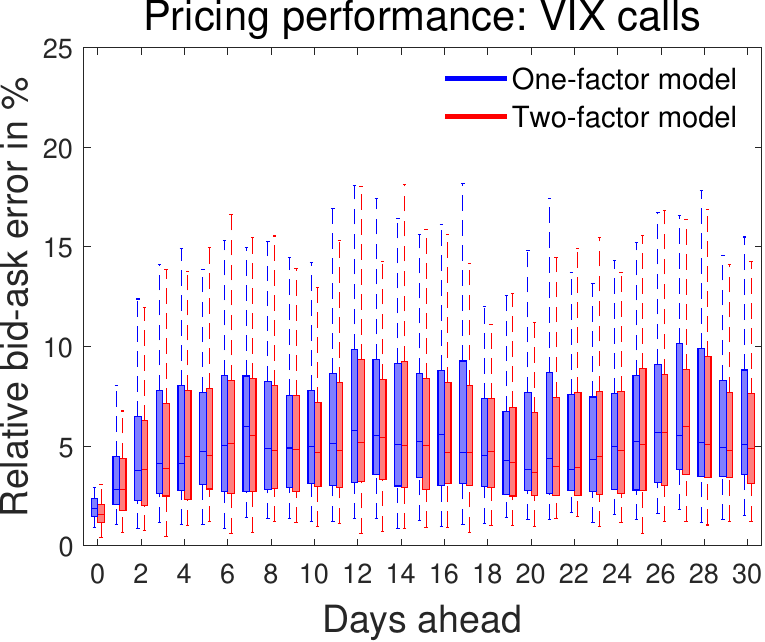}
    \end{minipage}
    \caption{Boxplots illustrating the daily average relative error distributions for one- and two-factor Bergomi models. The boxes display the $25$th, $50$th, and $75$th percentiles, while the whiskers extend to $1.5$ times the interquartile range from the $25$th and $75$th percentiles. \textit{Rows:} Tests 1 and 2 respectively. \textit{Columns:} VIX futures and calls.}
     \label{fig:boxplots-tests1-and-2} 
\end{figure}

\begin{figure}[h!]
    \centering
    \begin{minipage}{0.48\textwidth}
        \centering
        \includegraphics[width=\textwidth]{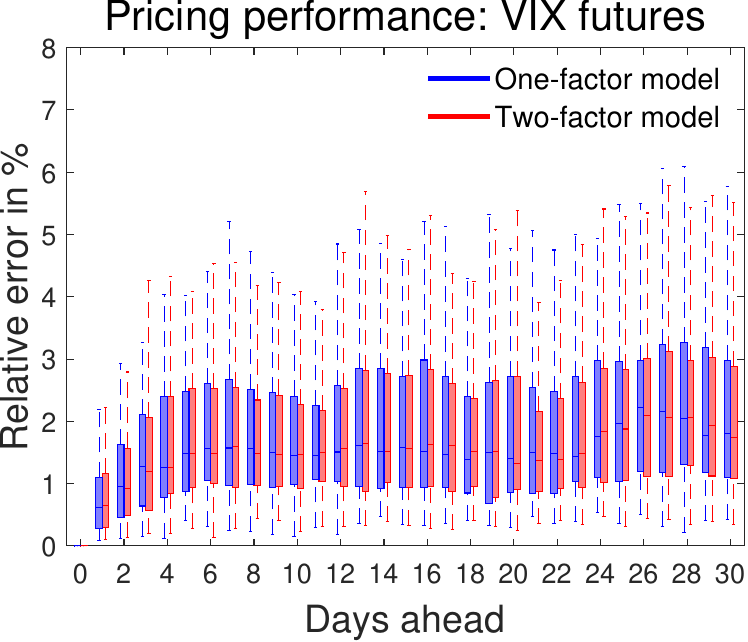}
    \end{minipage}
     \hspace{0.3cm}
    \begin{minipage}{0.48\textwidth}
        \centering
        \includegraphics[width=\textwidth]{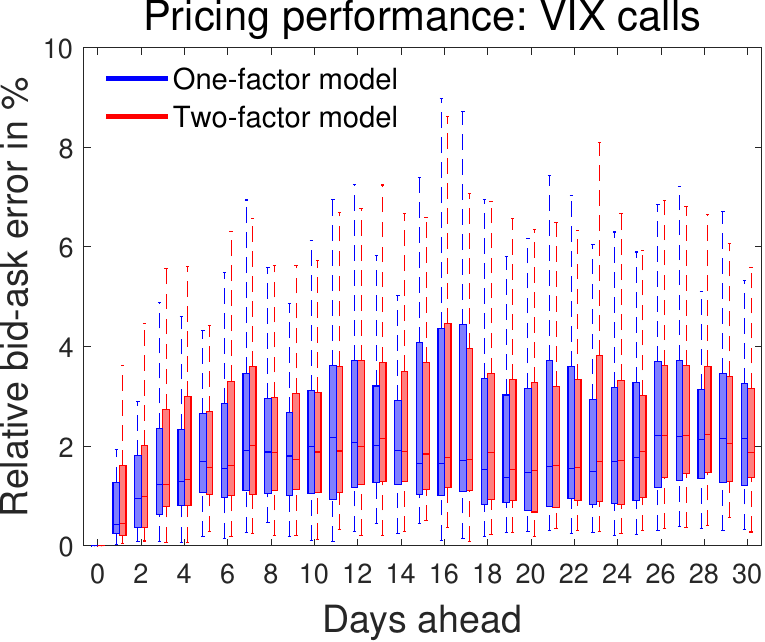}
    \end{minipage} \\
    \vspace{0.4cm}
        \begin{minipage}{0.48\textwidth}
        \centering
        \includegraphics[width=\textwidth]{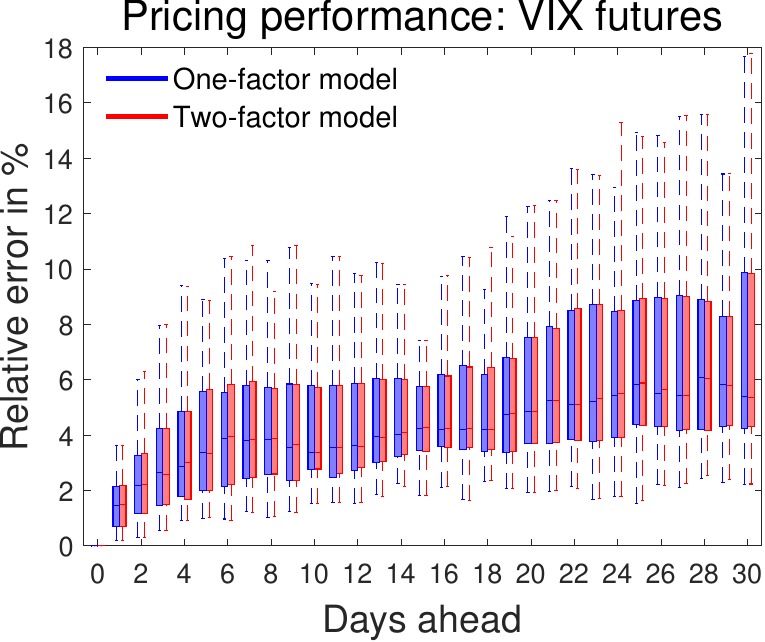}
    \end{minipage}
     \hspace{0.3cm}
    \begin{minipage}{0.48\textwidth}
        \centering
        \includegraphics[width=\textwidth]{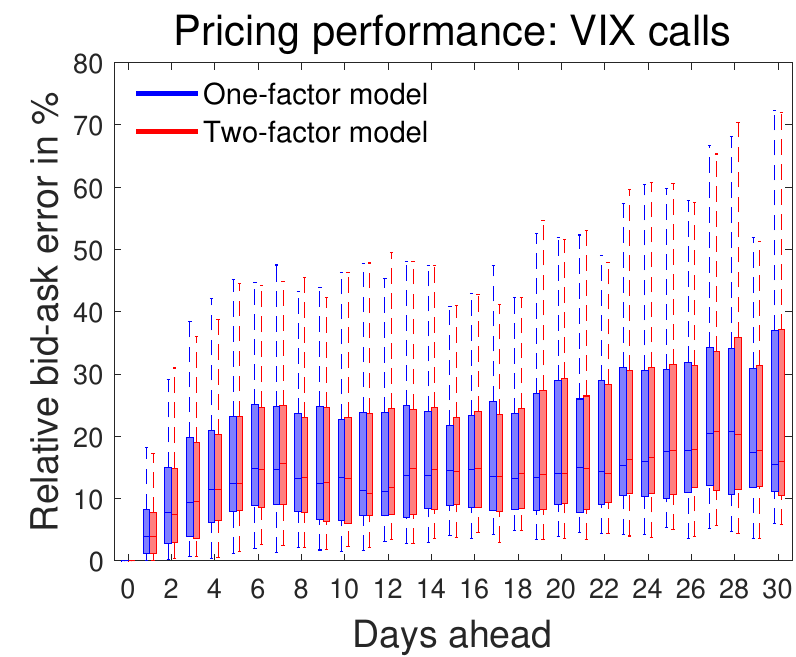}
    \end{minipage}
    \caption{Boxplots illustrating the daily average relative error distributions for one- and two-factor Bergomi models. The boxes display the $25$th, $50$th, and $75$th percentiles, while the whiskers extend to $1.5$ times the interquartile range from the $25$th and $75$th percentiles. \textit{Rows:} Tests 3 and 4 respectively. \textit{Columns:} VIX futures and calls.}
     \label{fig:boxplots-tests3-and-4} 
\end{figure}

Using~\eqref{eqn:are-futures} and~\eqref{eqn:arbae-calls}, for each test above,  we compute the daily average relative errors between the market and model prices for the global surface and display the distribution of the errors using the boxplots in Figures~\ref{fig:boxplots-tests1-and-2} and~\ref{fig:boxplots-tests3-and-4}. Tests 1 and 2 are illustrated in Figure~\ref{fig:boxplots-tests1-and-2} while Tests 3 and 4 appear in Figure~\ref{fig:boxplots-tests3-and-4}. Tests 1 and 2 are similar to the prediction of the SPX implied volatility surface done in \cite{abi2024volatility}, whereas Test 3 is similar to the test on model prediction quality using SPX options done in \cite{romer2022empirical}.

Unsurprisingly, the out-of-sample relative errors illustrated in the boxplots are larger than the in-sample relative errors of Section~\ref{sec:calibration}. Both models show similar accuracy in pricing, although the two-factor model has a slight edge over its one-factor counterpart in some cases, but clearly little is gained from using a two--factor model here, especially since each model has three calibratable parameters in its parametrisation. 
In the first three tests, $\xi_0^T$ is not held fixed, therefore providing a better fit to market prices. This is especially evident in Test 3, where $\xi_0^T$ is recalibrated daily based on the exact relationship between $\xi_0^T$ and model prices, rather than utilising the model-independent relationship \eqref{eqn:initial-variance-term-structure}. In fact, a relative pricing error on the order of one to two percent in the VIX futures can be attributed to this: In Test 2, this error is evident immediately, i.e., at zero days lag, while in Test 1 the VIX futures fit deteriorates to about this level within the first six days of recalculating $\xi_0^T$ using \eqref{eqn:initial-variance-term-structure}. Consequently, we can conclude that this is the main source of futures mispricing, i.e., holding the remaining calibrated parameters fixed has little impact on the quality of fit to the VIX futures market. For VIX call options, on the other hand, the fit is impacted by holding the parameters fixed. However, across all four of our tests we see that the relative bid/ask error stabilises. This means that to the fitting accuracy observed after about six days of holding parameters fixed, the Bergomi models considered here seem to reflect market dynamics well even out to 30 days (and most likely longer). Furthermore, also for VIX calls the primary driver of pricing accuracy is $\xi_0^T$, as can be seen when comparing Test 4 to the other three tests, and this term structure of forward variance can be calibrated to VIX futures (without recourse to market option prices).

It is worth noting that one of the differences between our tests and those of \cite{abi2024volatility, romer2022empirical} is that the tests of the latter focus on SPX options only, whereas ours involve VIX futures and options. This might be one of the reasons why our pricing errors are bigger than those in  \cite{abi2024volatility, romer2022empirical}.

\section{Conclusion}\label{sec:conclusion}
This study demonstrated the efficacy of quantisation for achieving fast and efficient calibration of the mixed one- and two-factor Bergomi models, with substantial gains in computational speed over the quadrature methods typically used in the prior literature. Our quantisation-based approach is twice as fast as exact quadrature in the one-factor model and approximately $120$ times faster in the two-factor model. This quantisation-based methodology allowed us to calibrate mixed Bergomi models to daily market prices for VIX futures and options, over a period covering several months. Remarkably, both models achieve near-perfect fits to VIX futures and calls in single--day cross--sectional calibration.  If the objective is solely to calibrate VIX futures and options, our results suggest that the mixed one-factor Bergomi model is sufficient. Both models exhibit reasonable stability of the calibrated parameters, as illustrated by our numerical tests for pricing under fixed parameters.  The calibration of the term structure of forward variance $\xi_0^T$ is the dominant driver of pricing accuracy for VIX futures in the mixed Bergomi models, with the remaining model parameters having little (separate) impact. For VIX call options, $\xi_0^T$ is still the primary driver of pricing accuracy, though the remaining model parameters have a more substantial impact on options than on futures, suggesting that if one extracts the term structure of forward variance from futures on a daily basis, substantially better option pricing performance is achieved even if the remaining parameters are not recalibrated. Interestingly, in all cases we considered, pricing errors when not recalibrating to the market stabilise after about six days of holding parameters fixed, rather than continuing to increase the further out we get from the calibration date. In this latter sense at least (i.e., to the accuracy of the pricing errors observed after about six days) one can say that the mixed Bergomi models describe VIX futures and option dynamics well.

\FloatBarrier  
\subsection*{Data and code availability statement:} MATLAB codes for quantisation accuracy tests are available on the GitHub page: 
\begin{center}
\href{https://github.com/nelsonkyakutwika/Mixed-Bergomi-VIX}{https://github.com/nelsonkyakutwika/Mixed-Bergomi-VIX}. 
\end{center}
For the empirical analysis, we use market data spanning $105$ days (five calendar months), obtained from the CBOE: \url{https://datashop.cboe.com/}.
\bibliographystyle{abbrv}
\bibliography{references}
\end{document}